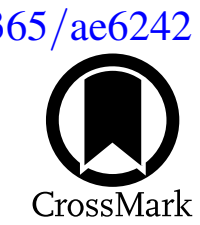

# A Value-added Physical Properties Catalog for Low-redshift Galaxies from DESI Legacy Imaging Surveys DR10

Shirui Wei[1,2,3], Changhua Li[1,3], Yanxia Zhang[1], Chenzhou Cui[1,3], Jinghang Shi[1,2], Wujun Shao[1,2,3], Zihan Kang[1,2], Yongheng Zhao[1], and Maoyuan Huang[1,3]

[1] National Astronomical Observatories, Chinese Academy of Sciences, Beijing 100101, People's Republic of China; lich@bao.ac.cn, zyx@bao.ac.cn, ccz@bao.ac.cn
[2] University of Chinese Academy of Sciences, Beijing 100049, People's Republic of China
[3] National Astronomical Data Center, Beijing 100101, People's Republic of China

## Abstract

Galaxy physical properties—such as star formation rate (SFR), stellar mass, and gas-phase metallicity—are essential for population studies and evolutionary analyses. Deriving these quantities for billions of galaxies in modern imaging surveys presents significant challenges due to limited spectroscopy and the computational costs associated with traditional spectral energy distribution fitting. As a result, many galaxies in large photometric surveys still lack homogeneous property estimates. This study introduces a multimodal deep learning model that integrates optical imaging with photometric catalog features to estimate SFR, stellar mass, and oxygen abundance in low-redshift galaxies. The model incorporates a ResNet-based convolutional neural network to extract spatial information from multiband images and a multilayer perceptron that processes catalog-level photometric features, leveraging complementary constraints from morphology, surface brightness, and broadband colors. Trained on reference measurements from the MPA-JHU DR8 catalog, the model is optimized for efficient large-scale estimation. When applied to the DESI Legacy Imaging Surveys (LS) DR10, the model generates a value-added catalog containing physical property estimates for approximately 547 million galaxies with redshifts $z \leqslant 0.5$. Validation through comparisons with independent catalogs and exploration of key scaling relations demonstrates that while the derived properties are not intended for precision measurements of individual objects, they effectively capture the dominant astrophysical trends necessary for ensemble studies. This catalog represents the first homogeneous set of photometry-based SFR, stellar mass, and metallicity estimates for DESI LS DR10, providing a vital resource for statistical studies of galaxies in the local Universe and facilitating comparisons with current and future spectroscopic surveys.

## 1. Introduction

Understanding the fundamental physical properties of galaxies—such as the star formation rate (SFR), stellar mass ($M_*$), and metallicity—is essential for studying galaxy formation and evolution. The SFR measures how efficiently a galaxy is turning gas into stars, while $M_*$ traces its integrated past star formation history (SFH). Metallicity, and in this work specifically, the gas-phase metallicity, defined as the abundance of heavy elements relative to hydrogen, reflects the distribution and enrichment of metals in the interstellar medium. These properties jointly describe a galaxy's stellar content, star-forming activity, and chemical composition, and they are closely connected through well-established scaling relations such as the star-forming main sequence (J. Brinchmann et al. 2004; D. Elbaz et al. 2007; G. Rodighiero et al. 2011; K. E. Whitaker et al. 2012; J. S. Speagle et al. 2014; C. Schreiber et al. 2015; A. Saintonge et al. 2016; N. M. Förster Schreiber & S. Wuyts 2020) and the mass–metallicity relation (MZR; J. Lequeux et al. 1979; C. A. Tremonti et al. 2004; L. J. Kewley & S. L. Ellison 2008; H. J. Zahid et al. 2014, 2017). Such relations help distinguish between star-forming and quiescent galaxies (G. Kauffmann et al. 2003; I. K. Baldry et al. 2004; K. G. Noeske et al. 2007) and reveal the effects of processes like gas inflow and outflow (C. A. Tremonti et al. 2004; S. J. Lilly et al. 2013; H. J. Zahid et al. 2014). Accurately measuring these properties for large galaxy samples is therefore crucial for building a picture of galaxy populations and understanding the physical mechanisms that shape them, ultimately providing insights into galaxy evolution models and, more broadly, the distribution of baryonic matter on a cosmic scale (M. R. Blanton & J. Moustakas 2009; P. Madau & M. Dickinson 2014; R. S. Somerville & R. Davé 2015; R. H. Wechsler & J. L. Tinker 2018).

Spectroscopic observations provide the most reliable constraints on galaxy physical properties. SFR and $M_*$ can be derived either from strong nebular emission-line modeling (e.g., G. Kauffmann et al. 2003; J. Brinchmann et al. 2004), or from full spectral fitting techniques that decompose a galaxy spectrum into mixtures of simple stellar populations of known ages, metallicities, and chemical enrichment (e.g., R. Cid Fernandes et al. 2005; M. Cappellari 2017). Gas-phase metallicity, in particular the oxygen abundance commonly expressed as 12 + log(O/H), can be estimated from strong nebular emission lines such as [O II], [O III], H$\beta$, and [N II] in star-forming H II regions (e.g., C. A. Tremonti et al. 2004), which trace the chemical enrichment of the ionized gas.

However, obtaining high-quality spectra for large galaxy samples remains observationally expensive. Even with ongoing large spectroscopic surveys such as the Dark Energy Spectroscopic Instrument (DESI; DESI Collaboration et al. 2016a) and Euclid (R. J. Laureijs et al. 2010), the number of galaxies with spectra is far smaller than the billions of sources detected in new-generation photometric imaging surveys, including the DESI Legacy Imaging Surveys (DESI LS; A. Dey et al. 2019) and the Legacy Survey of Space and Time (LSST; Ž. Ivezić et al. 2019). This severe mismatch between photometric and spectroscopic data volumes motivates the need for methods that can measure physical properties directly from photometric data.

The conventional approach for deriving galaxy physical properties from photometric data is spectral energy distribution (SED) fitting, which is conceptually similar to full spectrum fitting but applied to sparse photometric data points. Commonly used SED-fitting tools include CIGALE (M. Boquien et al. 2019), Prospector (B. D. Johnson et al. 2021), and BEAGLE (J. Chevallard & S. Charlot 2016). Despite its widespread use, SED fitting suffers from several well-known limitations (J. Walcher et al. 2011; C. Conroy 2013). The limited number of photometric data points leads to severe parameter degeneracies, most notably among stellar age, metallicity, and dust properties (G. Worthey 1994; C. Papovich et al. 2001; D. M. Wilkinson et al. 2017). In addition, SED fitting requires strong prior assumptions on the underlying models, including initial mass function (IMF), SFH, and dust attenuation laws (C. Conroy et al. 2009; A. C. Carnall et al. 2019). The derived physical properties can be highly sensitive to these modeling choices. For example, stellar mass estimates may differ by up to a factor of $\sim$2 (M. Siudek et al. 2024). Many SED-fitting algorithms rely on computationally expensive techniques such as Markov Chain Monte Carlo sampling, which makes them impractical for application to the vast galaxy samples produced by large-scale surveys (J. Woo et al. 2024). These limitations motivate the development of alternative, data-driven methods.

To overcome the limitations of traditional spectroscopic and SED-fitting methods, machine learning (ML) has emerged as a powerful alternative. ML models are capable of learning complex, nonlinear mappings between photometric data and galaxy physical properties with high computational efficiency, without requiring explicit parametric assumptions about SFHs or dust models. Once trained, ML models provide fast and reproducible estimation for large samples of galaxies, avoiding the need to refit models with different parameterizations for each object, making this approach well suited for new-generation large survey projects.

In recent years, ML-based approaches in this domain can be broadly divided into two categories. The first relies on multiband fluxes or magnitudes derived from photometric catalogs, which are used as inputs to regression models, including traditional ML algorithms such as CatBoost, Random Forests, and artificial neural networks (e.g., multi-layer perceptrons, MLPs), to estimate galaxy properties. These properties include photometric redshifts (e.g., M. J. Way & A. N. Srivastava 2006; A. Collister et al. 2007; S. Carliles et al. 2008; J.-T. Schindler et al. 2017; X. Jin et al. 2019; C. Li et al. 2022, 2024), as well as SFR, stellar mass, specific SFR, and metallicity (e.g., K. Stensbo-Smidt et al. 2017; G. Ucci et al. 2017; V. Bonjean et al. 2019; M. Delli Veneri et al. 2019; S. Surana et al. 2020; S. Mucesh et al. 2021; Euclid Collaboration et al. 2024b, 2025a, 2025b; F. Z. Zeraatgari et al. 2024). However, these approaches typically discard the rich spatial information available in imaging data.

With the development and application of convolutional neural networks (CNNs; Y. LeCun et al. 1989), several studies have explored the estimation of galaxy physical properties directly from images (e.g., B. Hoyle 2016; A. D'Isanto & K. L. Polsterer 2018; M. Ntampaka et al. 2019; J. Pasquet et al. 2019; J. F. Wu & S. Boada 2019; T. Buck & S. Wolf 2021; B. Dey et al. 2022; J. Chu et al. 2024; Euclid Collaboration et al. 2024a; M. Treyer et al. 2024; J. Zhong et al. 2024; J.-H. Cai et al. 2025; O. Torbaniuk et al. 2025; X. Zhou et al. 2025). Imaging data encode spatial information such as morphology and surface-brightness distribution, which are linked to galaxy properties: clumpy star-forming regions, spiral arms, and asymmetric structures are commonly associated with enhanced star formation activity (C. J. Conselice 2003; D. M. Elmegreen et al. 2005; H. M. Yesuf et al. 2021); global light distributions, such as concentration or bulge-to-disk ratio, are related to stellar mass-to-light ratios (E. F. Bell & R. S. de Jong 2001; G. Kauffmann et al. 2003); and disturbed morphologies can signal merger-driven gas inflows that affect chemical enrichment (L. Michel-Dansac et al. 2008; S. L. Ellison et al. 2013).

Photometric catalogs and imaging data therefore provide complementary views of galaxies. Combining these different data types can help break degeneracies present in single-data-type models and improve the robustness of galaxy physical property estimation. Recently, multimodal research, which is to combine different types of data (such as photometric images, spectra, and photometric catalog data) simultaneously, has started to be applied in galaxy physical properties measurements (e.g., W. Dobbels et al. 2019; B. Henghes et al. 2022; Euclid Collaboration et al. 2023, 2025c; L. Yao et al. 2023; L. Doorenbos et al. 2024; M. Gai et al. 2024; C. Hahn et al. 2024; G. Martínez-Solaeche et al. 2024; L. Parker et al. 2024, 2025; W. Roster et al. 2024; P. Li et al. 2025). These studies have demonstrated superior performance compared to methods based on a single data type, highlighting the potential of multimodal approaches for large photometric surveys.

In this paper, we propose a multimodal deep learning model that combines CNNs and MLPs to jointly exploit photometric imaging and catalog data, leveraging their complementary strengths to improve the estimation of the SFR, $M_*$, and $12 + \log(\mathrm{O/H})$. We apply our model to the DESI LS DR10, one of the largest multiband photometric catalogs currently available, and generate a catalog of these three physical properties. This physical properties catalog offers a valuable resource for statistical studies of galaxy populations and for direct comparisons with existing and future spectroscopic measurements from DESI and other surveys.

The structure of this paper is as follows. Section 2 describes the photometric datasets and reference samples of galaxy physical properties used for model training, validation, and catalog generation, including the DESI LS DR10 and the MPA-JHU DR8 catalog. Section 3 presents the architecture and training strategy of the multimodal deep learning model. Section 4 provides a detailed evaluation of the model performance. In Section 5, we apply the trained model to the DESI LS DR10 and construct the physical properties catalog. We then conduct a comprehensive analysis of its global properties, compare the results with existing measurements, and demonstrate potential scientific applications. Finally, Section 6 summarizes this work.

## 2. Data

### 2.1. DESI Legacy Imaging Surveys

DESI (M. Levi et al. 2013; DESI Collaboration et al. 2016a, 2016b) is a 5 yr spectroscopic redshift survey designed to map the 3D structure of the Universe over the redshift range $0 \leqslant z \lesssim 4$. By obtaining precise redshifts for nearly 40 millions of extragalactic sources, DESI aims to place unprecedented constraints on dark energy, large-scale structure, neutrino masses, and models of primordial inflation (DESI Collaboration et al. 2024a). In 2025 March, the DESI collaboration released its first public data release (DR1; DESI Collaboration et al. 2025), which includes all data acquired during the first 13 months of the main survey, together with a uniform reprocessing of data from the Survey Validation phase (DESI Collaboration et al. 2024a, 2024b). DR1 provides reliable redshift measurements for approximately 18.7 million unique targets over more than 9000 deg$^2$, including 13.1 million galaxies, 1.6 million quasars, and 4 million stars, making DESI the largest extragalactic spectroscopic redshift survey ever conducted to date.

DESI identifies its primary spectroscopic targets—Milky Way stars (MWS), Bright Galaxy Survey (BGS) galaxies, luminous red galaxies, emission-line galaxies, and quasars (QSOs)—using the DESI Legacy Imaging Surveys (DESI LS; A. Dey et al. 2019). The DESI LS comprises three individual surveys: the Dark Energy Camera (DECam) Legacy Survey (DECaLS; B. Flaugher et al. 2015), the Beijing-Arizona Sky Survey (H. Zou et al. 2017), and the Mayall $z$-band Legacy Survey. Together, these surveys provide deep optical imaging over approximately 14,000 deg$^2$ in the $g$, $r$, and $z$ bands and additionally augmented with four infrared observations in the W1 – W4 bands from the Near-Earth Object Wide-field Infrared Survey Explorer Reactivation Mission (A. Mainzer et al. 2014). The DESI LS DR10 is the tenth public data release of the DESI LS. Building upon the original footprint, DR10 further incorporates imaging data from several external DECam programs in the southern sky, including the Dark Energy Survey (Dark Energy Survey Collaboration et al. 2016), the DECam Local Volume Exploration Survey (A. Drlica-Wagner et al. 2021), and the DECam eROSITA Survey (A. Zenteno et al. 2025). This extension increases the total sky coverage to more than 20,000 deg$^2$ and provides an additional $i$ band in DECaLS.

DESI LS DR10 provides calibrated optical and infrared imaging data, from which photometric catalogs are constructed using TRACTOR (D. Lang et al. 2016). The TRACTOR measures source fluxes by fitting parametric surface-brightness models to the imaging data, including a point-spread function model for point sources and several extended galaxy models, such as round exponential (REX), de Vaucouleurs (DEV), exponential (EXP), and Sérsic (SER) profiles. Based on the best-fitting model, DESI LS DR10 delivers model-based total fluxes, together with aperture photometry measured within a set of eight radii (ranging from $0\rlap{.}''5$–$7\rlap{.}''0$) for optical bands and five radii for infrared bands (ranging from $3''$–$11''$). Notably, the infrared fluxes are obtained through forced photometry at the optical source positions. Galactic extinction values $E(B - V)$ are also included and used to correct for Galactic extinction.

In addition to the catalog-level photometry, DESI LS DR10 provides coadded images for each sky brick, combining multiple exposures to deliver deeper imaging suitable for image-based analyses. The optical images are projected onto a simple WCS TAN projection with a pixel scale of $0\rlap{.}''262$ pixel$^{-1}$, while the Wide-field Infrared Survey Explorer (WISE) images are provided with a pixel scale of $2\rlap{.}''75$ pixel$^{-1}$. These complementary catalog and imaging data are used in this work for the multimodal estimation of galaxy physical properties.

### 2.2. Reference Galaxy Physical Properties: MPA-JHU DR8

In this work, we employ the deep learning method to learn the mapping between photometric data and galaxy physical properties. This requires a reliable dataset that provides accurate physical properties used as labels for supervised training. We adopt the MPA-JHU DR8, a commonly used value-added catalog produced by a collaboration between the Max Planck Institute for Astrophysics and Johns Hopkins University (G. Kauffmann et al. 2003; J. Brinchmann et al. 2004), as the source of training labels. This catalog provides key measurements of galaxy physical properties (e.g., SFR, stellar mass, nebular oxygen abundance, and specific SFR) for 1,843,200 Sloan Digital Sky Survey (SDSS) DR8 galaxies up to $z \sim 0.7$. The MPA-JHU catalog has been extensively used as a benchmark for comparisons with alternative physical property measurements and galaxy formation simulations (e.g., Y.-Y. Chang et al. 2015; J. Schaye et al. 2015; M. Siudek et al. 2024). It has also enabled numerous studies of empirical relations among galaxy properties, such as the fundamental metallicity relation (e.g., F. Mannucci et al. 2010), as well as investigations of different galaxy evolution pathways (e.g., Y.-J. Peng et al. 2010). Moreover, the catalog has served as the reference source of training labels in several recent works that apply deep learning techniques to estimate galaxy physical properties (e.g., V. Bonjean et al. 2019; F. Z. Zeraatgari et al. 2024; J. Zhong et al. 2024; J.-H. Cai et al. 2025).

Since our focus is primarily on the SFR, $M_*$, and $12 + \log(\mathrm{O/H})$, we briefly summarize how the three physical properties are calculated in the MPA-JHU catalog. SFR is derived using a combination of spectroscopic emission lines within the SDSS fiber and broadband photometry with aperture corrections outside the fiber (J. Brinchmann et al. 2004; A. Gallazzi et al. 2005; S. Salim et al. 2007). For galaxies with weak emission lines or those classified as active galactic nuclei (AGNs), their SFRs are measured solely from photometric measurements. Stellar masses are estimated using the Bayesian methodology described in G. Kauffmann et al. (2003), based primarily on SDSS $ugriz$ photometry, with both fiber and model magnitudes corrected for nebular emission. We adopt the total stellar mass from the catalog based on model magnitudes. Gas-phase oxygen abundances are estimated using strong optical emission lines, including [O II] $\lambda$3727, H$\beta$, [O III] $\lambda$5007, [N II] $\lambda\lambda$6548, 6584, and [S II] $\lambda\lambda$6717, 6731 following the Bayesian method outlined in C. A. Tremonti et al. (2004) and J. Brinchmann et al. (2004). And the oxygen abundances are provided only for star-forming galaxies.

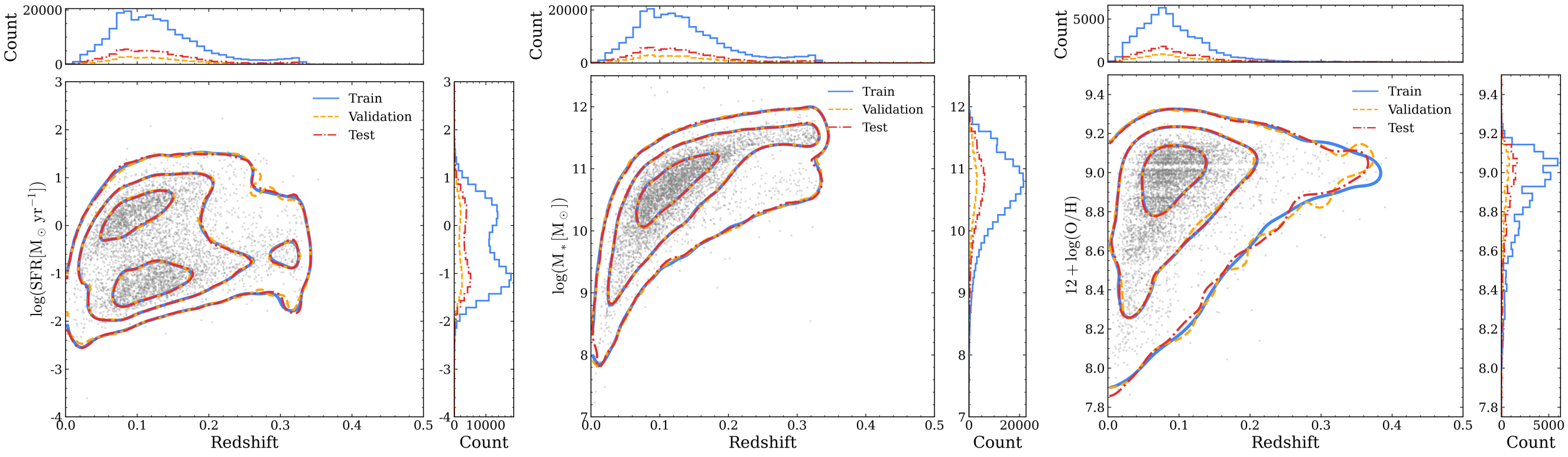

**Figure 1.** Distributions of SFR, stellar mass ($M_*$), and oxygen abundance (12 + log(O/H)) as a function of redshift for the training, validation, and test samples. Gray points represent a random subset of 5000 galaxies drawn from the full dataset, shown only to illustrate the overall density distribution. The contours are computed separately for each dataset (training, validation, and test), enclosing 39%, 86%, and 98% of the corresponding sample, which approximately correspond to the $1\sigma$, $2\sigma$, and $3\sigma$ levels of a 2D histogram. The top and right panels show the marginal distributions in redshift and in the corresponding physical properties, respectively. The blue solid, orange dashed, and red dotted–dashed lines denote the training, validation, and test sets, respectively.

**Table 1**
Summary of the Selection Criteria and the Meanings of the Relevant Flags Used in Constructing the Training Sample from the MPA-JHU DR8 Catalog

| Property/Flag | Selection | Description |
|---|---|---|
| RELIABLE | ≠0 | Indicates reliable derived physical properties; a value of 0 denotes unreliable measurements. |
| Z | >0 | Spectroscopic redshift. |
| Z_WARNING | =0 | Flag for redshift quality; nonzero values indicate unreliable redshift measurements. |
| TARGETTYPE | Galaxy | Object classified as a galaxy. |
| SFR_TOT_P50 | ≠−9999 | Median of the logarithmic total SFR posterior distribution; −9999 indicates missing values. |
| LGM_TOT_P50 | ≠−9999 | Median of the logarithmic total stellar mass posterior distribution; −9999 indicates missing values. |
| OH_P50 | ≠−9999 | Median of the oxygen abundance posterior distribution; −9999 indicates missing values. |

The MPA-JHU catalog provides these physical properties in logarithmic form, together with several percentile estimates that characterize their posterior probability distributions (2.5th, 16th, 50th, 84th, and 97.5th percentiles). Specifically, the median values (50th percentiles; denoted as SFR_TOT_P50, LGM_TOT_P50, and OH_P50 in the catalog) are adopted throughout this work as the reference values and are used as labels for model training.

To construct a reliable training sample, we apply a series of quality cuts to the MPA-JHU DR8 catalog based on catalog flags. The definitions of these flags and the corresponding selection criteria are summarized in Table 1. For each selected galaxy, we adopt the 50th percentile values of the three properties as training labels, together with the spectroscopic redshifts from the MPA-JHU catalog, which are retained for subsequent analysis. We emphasize that the MPA-JHU DR8 physical properties are not direct physical ground truth but model-dependent estimates derived under specific assumptions (e.g., IMF, dust law, metallicity calibration). Consequently, our model should be interpreted as learning to reproduce the MPA-JHU DR8 measurement system in a photometry-based manner, rather than estimating fundamental physical quantities from first principles. Oxygen abundance labels are available only for star-forming galaxies in MPA-JHU DR8. Therefore, metallicity estimates for quiescent or AGN-host galaxies in our catalog should be interpreted with caution, as they represent model extrapolations conditioned on photometric similarity.

### *2.3. Training Sample Construction*

For deep learning approaches, a well-defined dataset for model training, validation, and test is required to learn the mapping between photometric catalogs, imaging data, and physical properties. Since our goal is to build a reliable deep learning model to estimate physical properties of the DESI LS DR10 galaxies, the training dataset is constructed directly from the DESI LS DR10 to ensure consistency between the training data and the final application sample.

To ensure reliable photometry, we apply a set of selection criteria to the DESI LS DR10 sources. We first restrict the sample to objects classified as galaxy in the DESI LS DR10 catalog, corresponding to sources whose best-fitting TRACTOR models are REX, DEV, EXP, or SER. We then exclude objects fainter than the DESI $5\sigma$ detection limits, namely $g > 24.0$, $r > 23.4$, and $z > 22.5$ (A. Dey et al. 2019). In addition, sources with maskbits≠0 are removed, as this flag indicates potential issues such as saturated pixels, image artifacts, or contamination from nearby bright stars, globular clusters, or large galaxies.

The resulting galaxy sample is then cross-matched with the MPA-JHU DR8 catalog, which provides the reference physical properties using a positional matching radius of $1''$. The cross-matched results formed the final training sample. This sample contains 347,633 galaxies with SFR measurements, 373,023 galaxies with stellar mass measurements, and 76,762 star-forming galaxies with oxygen abundance estimates. The distributions of spectroscopic redshifts and galaxy physical properties are shown in Figure 1. Most galaxies are located at redshifts $z \leqslant 0.4$. The sample distributions of these three physical properties show a redshift dependence, with a steep rise at $z < 0.1$ that gradually flattens into a plateau at higher redshifts ($z > 0.1$). The dynamic ranges of the three properties differ significantly: the SFR spans approximately from

−3 to 2, the logarithm of stellar mass from 7 to 12, while the $12 + \log(O/H)$ covers a narrower range from 7.5 to 9.5.

### *2.4. Feature Construction*

To fully exploit the information contained in both the photometric catalogs and imaging data from DESI LS DR10, and to address issues such as missing values and Galactic extinction, we apply a series of preprocessing steps and construct a set of derived features for both modalities as inputs to the multimodal deep learning model.

For the photometric catalog data, we first compute AB magnitudes from the model and aperture fluxes provided by the DESI LS DR10 catalog. All magnitudes are corrected for Galactic extinction using the corresponding $E(B-V)$ values. Redshift is included as an explicit input feature, as galaxy physical properties evolve with cosmic time, and redshift encodes important observational effects relevant to their estimation.

As indicated in J. Zhong et al. (2024), color information traces the spectral features of galaxies and encodes information about their stellar populations and star formation activity, as it is particularly sensitive to variations in stellar age, metallicity, and dust attenuation, which govern galaxy physical properties. Therefore, incorporating color features alongside magnitudes provides more informative model inputs. Following the methodology of C. Li et al. (2024), who also constructed color features from DESI LS DR10 for photometric redshift estimation, we build a comprehensive set of candidate photometric features based on extinction-corrected magnitudes by computing magnitude differences. This results in a combined feature set of magnitudes and colors, yielding a total of 1626 features, including $E(B-V)$ and redshift. From this full feature set, we further select property-specific subsets to identify the most relevant features, remove redundant ones, and achieve a balance between computational efficiency and model performance. The selection is based on feature importance scores derived from CatBoost (A. Veronika Dorogush et al. 2018). CatBoost is an ML algorithm based on gradient boosting decision trees that can effectively capture nonlinear relationships and interactions among features, and has been widely applied in astronomical classification and regression tasks. Its feature importance is computed during the training process and reflects the contribution of each feature to the model prediction, enabling the identification of the most relevant inputs. This approach has been widely adopted for feature selection in related studies (e.g., F. Z. Zeraatgari et al. 2024). Through applying this algorithm, we select 28 features for SFR estimation, 50 features for stellar mass estimation, and all 1626 features for oxygen abundance estimation. The feature selection procedure and the complete lists of selected features are described in Appendix A.

Missing or invalid photometric values may arise from nondetections, negative fluxes due to background subtraction, or other processing issues. According to the DESI LS DR10 catalog description, there are many such cases for the faintest objects as expected. To handle such cases in a uniform way, we compute the maximum magnitude among all valid measurements in the sample, and use this value to fill missing entries for each band. This ensures that all feature values remain within a reasonable numerical range (rather than being infinite or NaN), which is beneficial for subsequent model training. Using the maximum magnitude also assigns these features a minimal contribution in the corresponding band. In Appendix B, we further examine the effect of filling missing values with maximum values on property estimation.

For the imaging data, we extract cutout images in the $g$, $r$, $i$, and $z$ bands centered on each target galaxy using its R.A. and decl. Each image is cropped to a fixed size of 64 × 64 pixels, which is sufficient to capture the background sky and the full extent of each galaxy while minimizing contamination from other objects. The native pixel scale of $0\farcs262\,\mathrm{pixel}^{-1}$ is preserved for all optical bands to maintain consistency with the source detection and photometric measurements. Although DESI LS DR10 also provides WISE images, we do not include them in the imaging inputs due to their pixel scale of $2\farcs75\,\mathrm{pixel}^{-1}$, which is inconsistent with the optical images and may dilute morphological information or introduce additional noise. But the WISE photometric measurements in the catalog are retained, as infrared emission provides complementary information to the optical data, particularly for tracing star formation activity and the underlying stellar populations (V. Bonjean et al. 2019).

Motivated by the importance of color information and inspired by similar studies (e.g., W. Roster et al. 2024; C. Zhang et al. 2024), we construct color images by computing pixel-wise differences between photometric bands. The final imaging input consists of seven channels in the following order: $g$, $r$, $i$, $z$, $g-r$, $r-i$, and $i-z$, resulting in an input tensor of shape (7, 64, 64) for each galaxy. For bands or channels with missing image data, the corresponding channel is filled with zeros to ensure a consistent input shape for all samples. By explicitly providing color information as additional input channels, this design alleviates the need for the model to infer such relationships implicitly from the flux images, which is inherently more challenging and may lead to less efficient feature learning. In Appendix A, we further examine the effectiveness of incorporating explicit color information into the image features and assess its impact on model performance.

In the end, each galaxy of the training sample is represented by a set of seven-channel optical images together with 28, 50 and 1626 photometric features for SFR, stellar mass, and oxygen abundance estimation, respectively. The optical images have a pixel scale of $0\farcs262$ per pixel, with each image covering a region of 64 × 64 pixels. These features jointly serve as the inputs to the multimodal deep learning model described in the following section.

## 3. Model

Building on our previous work (S. Wei et al. 2025), we develop a multimodal model that jointly leverages imaging and photometric data to improve the accuracy of galaxy physical property estimation. The overall architecture, illustrated in Figure 2, comprises three main components. The first component is a CNN based on the residual neural network (ResNet) series, which processes multiband optical images. The second component is an MLP that independently extracts information from catalog features. Features from both components are then concatenated and subsequently passed to the third component, another MLP, to estimate the target physical properties. In the following sections, we provide detailed descriptions of each component of the model.

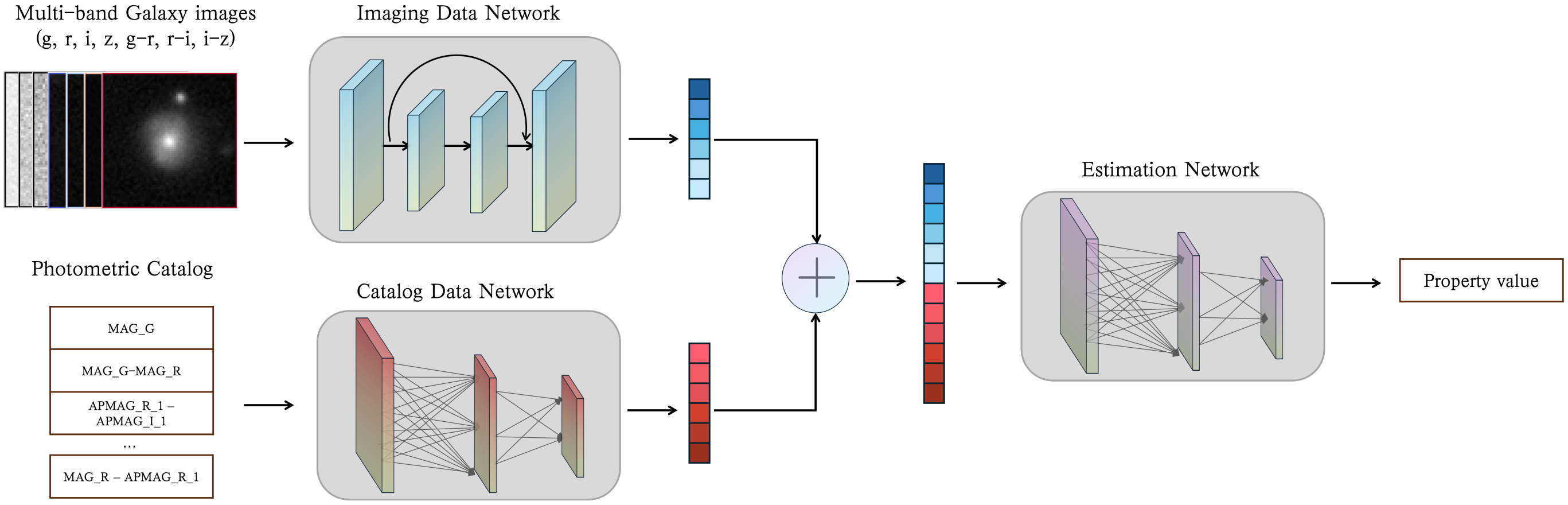

**Figure 2.** Schematic diagram of the multimodal model. The model consists of three components: an imaging data network that processes optical multiband images, a catalog data network that handles catalog features, and an estimation network. The image and catalog features are concatenated and fed into the estimation network to estimate galaxy physical properties.

### *3.1. Imaging Data Network*

Imaging data provide information on the apparent morphology of galaxies, including sizes, extended structures, and spatial distributions. To capture such features from multiband images for physical property estimation, we employ CNNs, which are well suited for image-processing tasks. Among the available architectures, we adopt the ResNet series, which is both computationally efficient and effective for extracting complex features from our multiband imaging data.

ResNet, proposed by K. He et al. (2016), is a deep convolutional architecture developed to address the degradation problem that commonly arises when training very deep neural networks. Its key innovation is the residual block (see the schematic diagram of Figure 2 in K. He et al. 2016), which consists of two convolutional layers together with a skip connection that directly links the block input to output. This design preserves the original information, alleviates vanishing gradients, and significantly stabilizes the training of deep models. Owing to this effective design, ResNet has become one of the most influential models in computer vision, and has also been successfully adopted in many scientific domains. In astronomy, ResNet-based methods have been applied to tasks such as astronomical object detection and classification (I. T. Andika et al. 2023), spectral feature identification (S. Kalvankar et al. 2021), and various image-processing applications (V. Gustafsson et al. 2024).

The overall ResNet architecture starts with a $7 \times 7$ convolutional layer followed by a $3 \times 3$ max-pooling layer to reduce spatial resolution and computational cost. The network then proceeds through a sequence of residual modules. At the beginning of each new module, the number of feature channels is doubled while the spatial resolution is halved, allowing the model to capture multiscale features while preserving the time complexity. Finally, the network ends with a global average pooling layer and a fully connected layer, which can be adapted for various downstream tasks.

Such architectures make ResNet well suited for extracting information from galaxy images. Galaxies often contain subtle or irregular structures—such as faint spiral arms and tidal tails, which require a deep network to capture effectively. The skip connections in ResNet enable stable training, allowing the model to learn these intricate patterns more thoroughly. In addition, galaxy features span multiple scales, from compact nuclear bulges to bars and spiral arms, and to extended outer halos. The hierarchical structure of ResNet provides multiscale receptive fields, facilitating the extraction of features across different spatial scales. Furthermore, compared with other models such as Visual Geometry Group network (K. Simonyan & A. Zisserman 2014), EfficientNet (M. Tan & Q. V. Le 2019), or Vision Transformer (A. Dosovitskiy et al. 2020), which often require larger datasets and long training time, ResNet provides a favorable balance between efficiency and performance for our task.

Based on the number of layers, which reflects the model scale, there are several ResNet variants, including ResNet-18, ResNet-34, ResNet-50, and ResNet-101. We adopt different variants for different physical properties predictions: ResNet-50 for SFR, ResNet-18 for stellar mass, and ResNet-34 for oxygen abundance, according to the results of our experiments. These choices reflect differences in the granularity and complexity of the information required for each property. Therefore, rather than using a single unified model for all properties, we select the variant that is most suitable for each specific task.

To adapt the model to our dataset, we modified several layers of the original ResNet architecture. Specifically, the input layer was adjusted to accommodate the seven channels of our multiband images, and the final fully connected layer was replaced with a 2048-dimensional layer, serving as a unified image representation rather than performing original classification.

### *3.2. Catalog Data Network*

Imaging data captures visual features such as shape and structure while catalog data provides precise numerical flux and color information. These two modalities encode complementary aspects of galaxy properties. To leverage the information contained in the catalog data, we concatenate the selected photometric features for each property into a 1D vector as the input, with the feature length varying across different properties. As shown in Figure 2, a separate MLP is employed to process these photometric feature vectors and extract the information independently. As a result, the model architecture can be viewed as a combination of two parallel components: a ResNet-based module for the imaging data channel and an MLP for the catalog data channel. This structure enables efficient feature extraction from each modality, while avoiding challenges associated with increased model complexity and interpretability.

### 3.3. Multimodal Fusion and Estimation Network

After being processed by the imaging and catalog data networks, the resulting features are concatenated and passed through an MLP to extract the fused representation for properties estimation.

The estimation of galaxy physical properties is treated as a regression problem, where the goal is to predict a single scalar value for each property (V. Bonjean et al. 2019; F. Z. Zeraatgari et al. 2024). In this work, we focus on point estimation and train the model by minimizing the mean squared error (MSE) between the predicted and reference values. The MSE loss is chosen based on empirical performance, as it provides stable convergence and is well suited for continuous physical quantities.

### 3.4. Model Training

The multimodal model is implemented using the PYTORCH framework (A. Paszke et al. 2019), and all experiments are performed on the China-VO Science Platform (C. Li et al. 2017) provided by the National Astronomical Data Center. The dataset is split into training, validation, and test sets with a ratio of 70:10:20. As shown in Figure 1, the target properties are consistent across the training, validation, and test sets, as indicated by the strong overlap of the contours. Given the large sky coverage and the relatively smooth variation of observational conditions across bricks, we expect potential spatial correlations between training and test samples to have a negligible impact on the reported performance metrics. In addition, to reduce the effect of galaxy orientation on our imaging data model, data augmentation is applied to the training set by performing random flips and rotations in 90° increments. We use the Weights & Biases sweep tool[4] to perform hyperparameter optimization on the validation set, obtaining an individually best-performing model for each target property.

## 4. Evaluation

In this section, we evaluate the performance of the proposed multimodal model on the test set to demonstrate its effectiveness in estimating galaxy physical properties.

### 4.1. Metrics

To quantify the performance of galaxy property estimation, we adopt several commonly used evaluation metrics in regression tasks, including MSE, root mean squared error (RMSE), normalized root mean squared error (NRMSE), mean absolute error (MAE), standard deviation ($\sigma$), bias, outlier fraction ($\eta$), coverage, normalized median absolute deviation (NMAD), and Pearson correlation coefficient ($r$).

We first define the residual for each sample as

$$\Delta y_i = y_{\mathrm{pred},i} - y_{\mathrm{reference},i}, \tag{1}$$

[4] https://wandb.ai/site/experiment-tracking/

where $y_{\mathrm{pred},i}$ denotes the predicted property value, and $y_{\mathrm{reference},i}$ is the corresponding reference value from the MPA-JHU DR8 catalog.

MSE is defined as

$$\mathrm{MSE} = \frac{1}{n}\sum_{i=1}^{n}(\Delta y_i)^2, \tag{2}$$

which measures the average squared deviation of predictions from the reference values.

RMSE is given by

$$\mathrm{RMSE} = \sqrt{\mathrm{MSE}}. \tag{3}$$

It represents the typical magnitude of the deviation from reference measurements in the same units as the target.

NRMSE is obtained by normalizing the RMSE with the dynamic range of the reference property values:

$$\mathrm{NRMSE} = \frac{\mathrm{RMSE}}{y_{\mathrm{reference,max}} - y_{\mathrm{reference,min}}}. \tag{4}$$

This provides a relative measure of the deviation from reference measurements, independent of the scale of the property.

MAE is defined as

$$\mathrm{MAE} = \frac{1}{n}\sum_{i=1}^{n}|\Delta y_i|. \tag{5}$$

It quantifies the average absolute deviation between predictions and reference values, giving a more robust sense of typical error.

The standard deviation ($\sigma$) is calculated as

$$\sigma = \sqrt{\frac{1}{n}\sum_{i=1}^{n}(\Delta y_i - \overline{\Delta y})^2}. \tag{6}$$

It captures the spread of residuals around the mean, reflecting the consistency of the predictions.

Bias is defined as the mean residual:

$$\mathrm{Bias} = \frac{1}{n}\sum_{i=1}^{n}\Delta y_i. \tag{7}$$

Bias indicates whether the model systematically overestimates or underestimates the property.

The outlier fraction $\eta$ is defined as the fraction of samples whose absolute residuals exceed 3 times the standard deviation:

$$\eta = \frac{1}{N}\sum_{i=1}^{N}\mathbb{I}(|\Delta y_i| > 3\sigma). \tag{8}$$

In addition, a coverage metric is defined as the fraction of predictions that fall within the uncertainty interval provided by the MPA-JHU DR8 catalog, characterized by the 16th and 84th percentiles of the posterior distributions. This metric serves as a consistency check between the predicted values and the catalog-derived uncertainty estimates:

$$\mathrm{Coverage} = \frac{1}{N}\sum_{i=1}^{N}\mathbb{I}(P_{16,i} \leqslant y_{\mathrm{pred},i} \leqslant P_{84,i}), \tag{9}$$

where $P_{16,i}$ and $P_{84,i}$ denote the 16th and 84th percentiles for each sample.

**Table 2**
Performance Comparison of the Proposed Multimodal Model and Its Single-modality Submodels

| Property | Modality | MSE | RMSE | NRMSE | MAE | $\sigma$ | Bias | Outlier | Coverage | NMAD | $r$ |
|---|---|---|---|---|---|---|---|---|---|---|---|
| SFR | Image | 0.174 | 0.418 | 0.053 | 0.304 | 0.416 | −0.032 | 0.012 | 0.799 | 0.338 | 0.848 |
| SFR | Catalog | 0.117 | 0.342 | 0.043 | 0.243 | 0.341 | **−0.016** | 0.012 | 0.879 | 0.262 | 0.901 |
| SFR | Multimodality | **0.115** | **0.339** | **0.042** | **0.240** | **0.338** | 0.025 | **0.011** | **0.891** | **0.261** | **0.903** |
| $M_*$ | Image | 0.033 | 0.181 | 0.028 | 0.116 | 0.180 | −0.019 | 0.015 | 0.647 | 0.110 | 0.958 |
| $M_*$ | Catalog | 0.018 | 0.134 | 0.021 | 0.069 | 0.134 | **0.002** | 0.017 | 0.871 | 0.060 | 0.976 |
| $M_*$ | Multimodality | **0.017** | **0.129** | **0.020** | **0.062** | **0.129** | **−0.002** | **0.016** | **0.898** | **0.051** | **0.978** |
| 12 + log(O/H) | Image | 0.010 | **0.010** | 0.061 | 0.071 | 0.099 | −0.014 | 0.016 | 0.393 | 0.077 | 0.893 |
| 12 + log(O/H) | Catalog | **0.008** | 0.091 | **0.055** | 0.066 | 0.090 | **0.005** | **0.013** | 0.401 | 0.073 | 0.911 |
| 12 + log(O/H) | Multimodality | **0.008** | 0.088 | **0.055** | **0.064** | **0.088** | **−0.005** | 0.014 | **0.416** | **0.072** | **0.918** |

**Note.** The best result for each metric is highlighted in bold.

NMAD is defined as

$$\mathrm{NMAD} = 1.4826 \times \mathrm{median}(|\Delta y_i|). \quad (10)$$

NMAD is approximately equivalent to the $\sigma$, with a reduced impact from extremely outlying errors.

$r$ is defined as

$$r = \frac{\sum_i (y_{\mathrm{pred},i} - \bar{y}_{\mathrm{pred}})(y_{\mathrm{reference},i} - \bar{y}_{\mathrm{reference}})}{\sqrt{\sum_i (y_{\mathrm{pred},i} - \bar{y}_{\mathrm{pred}})^2 \sum_i (y_{\mathrm{reference},i} - \bar{y}_{\mathrm{reference}})^2}}, \quad (11)$$

$r$ can quantify the strength of the linear correlation between the predicted and reference values.

## 4.2. Model Performance

### 4.2.1. Multimodal and Single-modality Model Comparison

To highlight the superiority of our multimodal model, we construct two single-modality baseline models by removing either the imaging data network or catalog data network: one retains only the MLP model for processing catalog data, and the other retains only the ResNet for imaging data. Their performances are then compared with the full multimodal model on the same dataset. Table 2 presents a quantitative comparison between the multimodal model and two single-modality baselines on the test set. Overall, the multimodal model consistently achieves the best or comparable performance across all three physical properties.

On one hand, the image-only model demonstrates non-negligible performance across all three properties, indicating that imaging data encodes physically meaningful information. In particular, galaxy morphology and spatial structure are physically correlated with star formation activity and stellar mass assembly, providing constraints that are not fully captured by integrated photometry alone.

On the other hand, the catalog-only model generally outperforms the image-only model. This is expected, as catalog data consist of aperture-matched and extinction-corrected photometry, whose fluxes and colors are more directly related to stellar population properties and are closely tied to traditional SED-based methods. In contrast, raw imaging data may suffer from contamination by background or nearby sources, which can introduce additional uncertainties.

Compared to single-modality models, the multimodal model demonstrates significant improvements, highlighting that imaging and catalog data offer complementary and nonredundant information. Their fusion effectively reduces degeneracies inherent in single-modality approaches, resulting in more accurate and robust estimation.

### 4.2.2. Property-dependent Performance

We analyze the model performance for each target property by comparing the multimodal predictions with the MPA-JHU reference values in Figure 3. All three properties show good agreement with the reference values, with most samples lying within the $\pm 3\sigma$ range. No significant redshift-dependent bias is observed, as the median residuals remain close to zero over the full redshift range. A mild positive offset is seen at very low redshift for all three properties, while oxygen abundance exhibits a systematic underestimation across the full redshift range. These trends are likely driven by the nonuniform and truncated data distribution, as shown in Figure 1, which leads to a slight regression toward the mean effect in the model predictions. This also reflects the varying levels of difficulty in estimating different physical properties. We evaluate the predictive performance of our model using both NRMSE and the empirical coverage of the predicted uncertainty intervals. Our results show that stellar mass and SFR are recovered with relatively low NRMSE and greater coverage, while gas-phase metallicity exhibits higher NRMSE and poorer coverage. This behavior is expected given the inherently more challenging nature of metallicity prediction, which leads to a discrepancy with respect to the uncertainty range provided by the MPA-JHU catalog. Unlike stellar mass and SFR, which can be reasonably constrained by broadband photometry, gas-phase metallicity estimation typically requires high-quality spectroscopic measurements of multiple nebular emission lines (e.g., C. A. Tremonti et al. 2004). Such detailed spectral data are not available in standard broadband photometric or imaging surveys, making metallicity prediction from these data alone a more indirect and complex process.

The mapping from photometric observables to metallicity is therefore highly degenerate. Different combinations of ionization conditions, star formation activity, and dust attenuation can produce similar photometric signatures while corresponding to distinct metallicities. This degeneracy limits the model's informational content and increases prediction scatter, as reflected in the higher NRMSE. Furthermore, intrinsic uncertainties in the metallicity labels—arising from variations in models and their underlying assumptions (C. A. Tremonti et al. 2004)—further constrain the achievable accuracy and

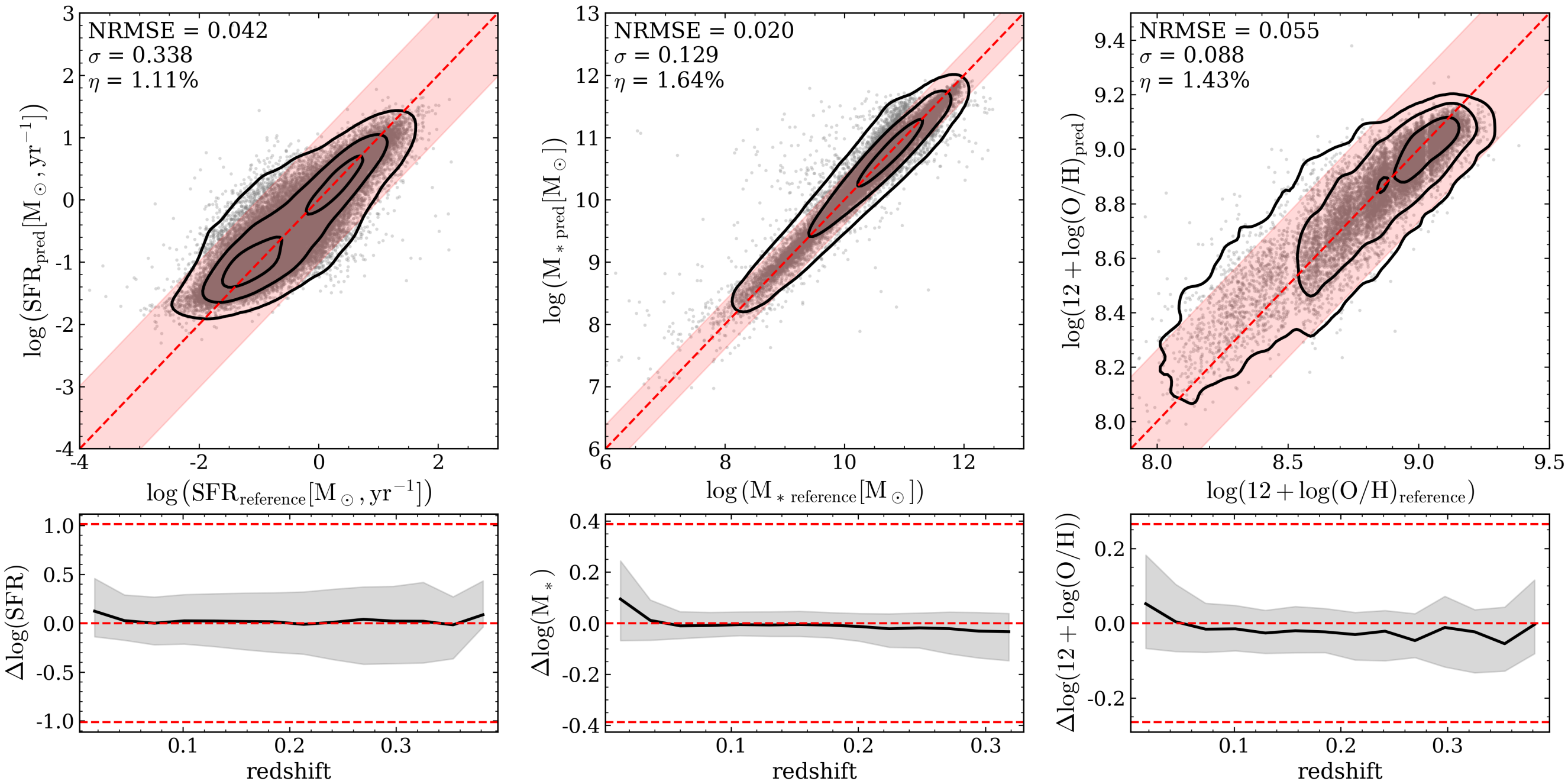

**Figure 3.** Comparison of star formation rate (SFR), stellar mass ($M_*$), and oxygen abundance (12+log(O/H)) between the multimodal model predictions and the MPA-JHU reference values for the test sample. Top panels: the gray points represent the test sample, while the black contours contain 39%, 86%, and 98% of the sample, corresponding to the $1\sigma$, $2\sigma$, and $3\sigma$ levels for a 2D histogram. The red dashed lines indicate the one-to-one relation, and the shaded regions mark the $\pm3\sigma$ range, beyond which predictions are considered outliers. Bottom panels: residuals ($\Delta$) as a function of redshift. The black solid line shows the median value in each redshift bin (with a typical bin width of $\sim$0.02), and the gray shaded region indicates the 16th–84th percentile range. The red dashed lines indicate $\Delta = 0$ and $\Delta = \pm3\sigma$, respectively.

**Table 3**
Model Performance Comparison with Related Studies

| Property | References | Method | Data | No. | RMSE | NRMSE | MAE | $\sigma$ | Bias | Outlier | NMAD |
|---|---|---|---|---|---|---|---|---|---|---|---|
| SFR | Our work | Our model | Multimodality | 347,633 | 0.337 | 0.042 | **0.238** | **0.336** | 0.024 | **0.011** | 0.258 |
| SFR | Bonjean+19 | RF | Catalog | 114,716 | ⋯ | ⋯ | ⋯ | 0.380 | ⋯ | ⋯ | ⋯ |
| SFR | Zeraatgari+24 | CatBoost | Catalog | 101,839 | **0.336** | **0.038** | 0.250 | ⋯ | **0.001** | **0.011** | ⋯ |
| $M_*$ | Our work | Our model | Multimodality | 373,023 | **0.122** | **0.019** | **0.061** | **0.122** | −0.003 | 0.018 | **0.051** |
| $M_*$ | Bonjean+19 | RF | Catalog | 114,716 | ⋯ | ⋯ | ⋯ | 0.160 | ⋯ | ⋯ | ⋯ |
| $M_*$ | Zeraatgari+24 | CatBoost | Catalog | 101,839 | 0.206 | 0.033 | 0.148 | ⋯ | **0.001** | **0.012** | ⋯ |
| $M_*$ | Zhong+24 | CNN | Image | 30,000 | ⋯ | ⋯ | 0.147 | 0.218 | ⋯ | ⋯ | ⋯ |
| $M_*$ | Siudek+24 | SED fitting | Catalog | 18,778 | ⋯ | ⋯ | ⋯ | ⋯ | ⋯ | ⋯ | 0.126 |
| 12+log(O/H) | Our work | Our model | Multimodality | 76,762 | **0.086** | **0.050** | **0.063** | **0.086** | −0.005 | 0.014 | 0.070 |
| 12+log(O/H) | Zeraatgari+24 | CatBoost | Catalog | 25,870 | 0.097 | 0.064 | 0.073 | ⋯ | **−0.000** | 0.011 | ⋯ |
| 12+log(O/H) | Cai+25 | CNN | Image | 5000 | ⋯ | ⋯ | ⋯ | 0.118 | 0.034 | **0.004** | ⋯ |

**Note.** No. indicates the test sample size reported in the literature; our model is re-evaluated on the full MPA-JHU DR8 dataset. "⋯" denotes metrics not reported. The best result for each metric is highlighted in bold. RF is short for random forests.

**References.** Bonjean+19 (V. Bonjean et al. 2019); Zeraatgari+24 (F. Z. Zeraatgari et al. 2024); Zhong+24 (J. Zhong et al. 2024); Siudek+24 (M. Siudek et al. 2024); Cai+25 (J.-H. Cai et al. 2025).

contribute to the apparent discrepancy between predicted uncertainties and empirical coverage.

Taken together, the higher NRMSE and poorer coverage for metallicity should not be interpreted as a failure of the model, but rather as a manifestation of the limited information content and intrinsic degeneracy of the input data with respect to gas-phase chemical abundances. Improving metallicity predictions will likely require incorporating observables more directly sensitive to emission lines, such as narrowband photometry or spectroscopic information, or adopting modeling frameworks that explicitly account for these degeneracies. This suggests that uncertainty calibration, particularly for metallicity, remains an important direction for future improvement.

#### *4.2.3. Comparison to Other Works*

To place our method in the context of existing studies, Table 3 presents a comparison between our multimodal model and several previous studies on galaxy physical property estimation that also use the MPA-JHU DR8 dataset for validation (V. Bonjean et al. 2019; M. Siudek et al. 2024; F. Z. Zeraatgari et al. 2024; J. Zhong et al. 2024;

J.-H. Cai et al. 2025). We provide a broad comparison by directly reporting the metrics from the literature, rather than enforcing evaluation to the same test samples. Our model is re-evaluated on the full MPA-JHU DR8 dataset, resulting in a substantially larger and more diverse evaluation sample.

Overall, our multimodal model achieves competitive performance across all three physical properties, often yielding the best results among existing studies based on a single data modality. The consistency on the full MPA-JHU dataset demonstrates that our model exhibits good generalization ability rather than performing well only on the test sample. For SFR estimation, F. Z. Zeraatgari et al. (2024) reported slightly better performance in several metrics. This difference is likely attributable to the inclusion of additional wavelength information in their input data, as their model incorporates WISE infrared bands together with the SDSS $u$, $g$, $r$, $i$, $z$ bands, where the $u$ band in particular provides additional constraints on recent star formation activity. Despite this, our results remain comparable. These results further demonstrate that integrating complementary information from multiple data modalities offers clear advantages over single-modality approaches for reliable galaxy property estimation in large-scale surveys.

## 5. Application: Galaxy Physical Properties Estimation for DESI LS DR10

In the previous section, we validated the performance of our multimodal model, demonstrating its effectiveness in estimating galaxy physical properties. This section describes the preparation of the full multimodal dataset from DESI LS DR10, the application of the trained model for large-scale estimation, an overview of the resulting catalog, and its scientific validation and applications.

### *5.1. Data Preparation and Estimation Pipeline*

We construct the multimodal dataset based on the DESI LS DR10 data products. In DESI LS DR10, coadded images and TRACTOR catalogs are organized hierarchically by R.A., with top-level directories indexed from 000 to 359 to facilitate large-scale storage and processing. Within each R.A. group, the data products are further divided into "bricks" of an approximate size of $0.25° \times 0.25°$. Each brick is defined as a box in R.A. and decl. coordinates. There are approximately 2.8 billion unique sources in the full catalog, distributed over 366,898 unique bricks. The information in the catalog follows the description in Section 2. To uniquely identify each source across all catalog entries, image cutouts, and the final catalog, we construct a unique identifier, `desiid`, by combining the `release`, `brickid`, and `objid` columns provided in the TRACTOR catalog.

For the image data, we generate multiband cutout images for each source using its sky coordinates. The cutout size and pixel resolution are kept identical to those used in the training dataset to ensure consistency between training and subsequent application. For the catalog data, we also apply the same preprocessing steps, including AB magnitude conversion, extinction correction, missing-value handling, and the computation of photometric features for each target physical property. The processed catalog features are then matched to their corresponding image cutouts using the `desiid` identifier.

Here we emphasize the redshift information adopted in this work. Redshift information plays a crucial role in galaxy property estimation. However, the raw LS DR10 catalog has not provided redshift values. For the training dataset, we adopt spectroscopic redshifts from the MPA-JHU DR8 catalog, which covers only a limited subset of galaxies. By contrast, the full LS DR10 sample requires a substantially larger redshift dataset.

We first collect spectroscopic redshifts from a wide range of public spectroscopic surveys, including DESI DR1, the Two Degree Field Galaxy Redshift Survey (M. Colless et al. 2003), the Six Degree Field Galaxy Survey (D. H. Jones et al. 2009), the Complete Calibration of the Color-Redshift Relation survey (D. C. Masters et al. 2017), the VIMOS VLT Deep Survey (O. Le Fèvre et al. 2013), zCOSMOS (S. J. Lilly et al. 2009), the DEEP2 Galaxy Redshift Survey (J. A. Newman et al. 2013), the PRIsm MUlti-object Survey (A. L. Coil et al. 2011; R. J. Cool et al. 2013), the Fiber Multi-Object Spectrograph COSMOS survey (D. Kashino et al. 2019), the VIMOS Public Extragalactic Redshift Survey (M. Scodeggio et al. 2018), the Large Sky Area Multi-Object Fiber Spectroscopic Telescope survey Data Release 8 (X.-Q. Cui et al. 2012; A. L. Luo et al. 2015), the WiggleZ Dark Energy Survey (M. J. Drinkwater et al. 2010), the Galaxy And Mass Assembly survey Data Release 3 (J. Liske et al. 2015), and the Optical Redshifts for the Dark Energy Survey (C. Lidman et al. 2020). For each survey, we filter out low-quality redshift measurements according to the quality criteria provided in the respective catalogs. We then cross-match the compiled spectroscopic redshift sample with the full LS DR10 catalog, resulting in 11,585,101 galaxies with reliable spectroscopic redshifts.

However, sources with spectroscopic redshifts constitute only a small fraction of the full LS DR10 catalog. To improve redshift coverage, we supplement spectroscopic measurements with photometric redshifts. We adopt the photometric redshift catalog presented by C. Li et al. (2024), which provides redshift estimates for approximately 1.5 billion galaxies, using a combination of the template-fitting code EAZY (G. B. Brammer et al. 2008) and the ML method CatBoost. For the DESI LS DR10 subset, this catalog achieves an NMAD scatter of 0.0098 and an outlier fraction of 0.08% (see Table 11 of C. Li et al. 2024). We cross-match the LS DR10 catalog with this photometric redshift catalog to assign photometric redshifts where available. Sources lacking both spectroscopic and photometric redshift estimates are excluded from the final sample. For the remaining sources, spectroscopic redshifts are adopted and used in subsequent photometric feature construction when available; otherwise, photometric redshifts are used instead. In Appendix B, we examine the effect of using photometric redshifts as a replacement on the model performance. The results indicate that physical properties estimates based on photometric redshifts can achieve overall statistical accuracy comparable to those based on spectroscopic redshifts.

Given that the redshifts of the training and validation samples are $<0.5$, with the majority below 0.3, predictions beyond the redshift range of the training sample are not expected to be reliable. We therefore restrict our analysis to DESI LS DR10 galaxies with $z \leqslant 0.5$ when applying the trained model. We caution against applying the model beyond the redshift range of the training sample ($z \leqslant 0.5$), as extrapolation

**Table 4**
Statistical Summary of the Catalog for the Full Galaxy Sample and the Subsample ($z \leqslant 0.3$)

| Sample | Total No. | Clean Photometry | Band Missing | Image Unavailable | Incomplete Cutouts |
|---|---|---|---|---|---|
| All galaxies | 547,656,694 | 90.1% | 89.8% | 16.6% | 0.70% |
| $z \leqslant 0.3$ | 130,863,053 | 89.6% | 88.3% | 13.2% | 0.86% |

**Note.** Percentages are computed relative to the total number of galaxies in each sample. "Clean photometry" refers to sources with maskbits = 0, indicating that the photometric measurements are not affected by known artifacts. "Band missing" indicates that at least one photometric band is unavailable in the catalog. "Image unavailable" denotes sources lacking cutout images in at least one band. "Incomplete cutouts" refer to sources whose image cutouts are partially missing or padded due to boundary effects.

in redshift space may lead to unphysical estimates. We implement a pipeline for large-scale estimation of galaxy physical properties across the survey. Details of the pipeline implementation will be presented in Wei et al. (2026, in preparation). Finally, the galaxy physical properties catalog was constructed and released.

### 5.2. Overview of the Physical Properties Catalog

In this section, we present a statistical overview of the galaxy physical properties catalog. This catalog contains 547,656,694 sources in 360 files, covering a redshift range up to $z = 0.5$. Table 4 summarizes the statistics of the catalog. Results for the subsample with $z \leqslant 0.3$, comprising the majority of the training set, are shown for reference.

An example catalog is shown in Table 5, which includes the unique object identifier, sky coordinates, redshift information, photometric magnitudes, and the three derived physical properties. In addition, several flag columns are provided to indicate the validity and quality of the estimated properties. A complete description of all catalog columns is summarized in Table 6. The catalog is publicly available at DOI: 10.12149/101777.

We first assess the reliability and applicability of the catalog using those flag columns. These flags enable users to construct customized selections for specific scientific goals.

1. maskbits: A bitwise mask from DESI LS DR10 catalog indicating whether an object is affected by potential photometric contamination. Sources with maskbits=0 are considered free from these issues. In the full catalog, 493,703,410 sources (90.1%) satisfy this criterion.
2. Cat_FLAG: This flag indicates the availability of valid photometric measurements across different bands. Sources with zero or negative raw fluxes in specific bands, which are filled with maximum values during preprocessing, are flagged as invalid (bit value 0), while valid measurements are marked as 1. The flag is represented as an eight-character string, where each position corresponds to one band ($g$, $r$, $i$, $z$, and W1 – W4).
3. Image_FLAG and Image_Padded: These flags describe the availability and integrity of image cutouts. If a corresponding band image is unavailable, the Image_FLAG is set to 0. For sources located near brick boundaries, where a complete cutout cannot be obtained, the missing regions are padded with zero values and flagged with a bit value of 2. And fully valid cutouts are assigned a value of 1. The flag is represented as an four-character string, where each position corresponds to one band ($g$, $r$, $i$, and $z$). The fraction of padded pixels relative to the total number of pixels in each cutout is recorded as Image_Padded. Overall, about 16.6% of sources lack corresponding image files in at least one band, while only 0.7% exhibit incomplete cutouts. Among them, the mean padding fraction is approximately 23%. Further improvements to the cutout generation algorithm will be addressed in future work.

To evaluate the impact of filled photometric values and padded image regions on the derived physical properties, we conduct a series of statistical analyses and controlled experiments presented in Appendix B, which may serve as a practical reference for assessing the reliability and usability of the catalog under different levels of data incompleteness.

Here, we apply only a minimal quality cut, using maskbits=0 to define a clean photometric sample, and examine their overall distributions. The distributions of redshift and the three physical properties are shown in Figure 4. For the three physical properties, the distributions of the galaxies in our catalog are broadly consistent with those reported in the reference MPA-JHU DR8 dataset (Figure 1) within the overlapping parameter space, showing no significant deviations beyond the expected range. In addition, the distributions of the $z \leqslant 0.3$ subsample are consistent with those of the full sample.

### 5.3. Scientific Validation and Applications

In this section, we compare our results with other existing VACs to examine their consistency, and explore scaling relations among the derived physical properties to demonstrate the potential scientific applications of our catalog.

To provide a clean and reliable baseline for visualization and comparison, we apply a strict quality selection. Specifically, we require maskbits = 0, CAT_FLAG = 11111111, Image_FLAG = 1111, and Image_Padded = 0. In addition, we select galaxies with available spectroscopic redshift measurements to ensure robust redshift estimates for the subsequent analysis. This selection yields a final "gold sample" consisting of 1,460,616 sources.

#### 5.3.1. Comparison with Other Catalogs

We perform a direct cross-catalog comparison between our catalog and several widely used galaxy properties catalogs, aiming to quantify systematic differences and validate the reliability of our estimates in the low-redshift region. The reference catalogs considered in this comparison include: (i) SDSS MPA-JHU DR8 (G. Kauffmann et al. 2003; J. Brinchmann et al. 2004; C. A. Tremonti et al. 2004), which serves as the sample for model training; (ii) eBOSS Firefly value-added catalog, ELODIE library (P. Prugniel et al. 2007) version, hereafter SDSS Firefly (J. Comparat et al. 2017);

**Table 5**
Example Entries from Our Physical Properties Catalog

| desiid | R.A. | Decl. | $z_{spec}$ | $z_{phot}$ | maskbits | MAG_R | log(SFR) | log($M_*$) | 12 + log(O/H) | Cat_FLAG | Image_FLAG | Image_Padded |
|---|---|---|---|---|---|---|---|---|---|---|---|---|
| 10000-155364-1056 | 0.640 | −31.915 | 0.108 | 0.157 | 0 | 17.166 | 0.584 | 10.438 | 9.051 | 11111111 | 1111 | 0.000 |
| 10000-155364-1057 | 0.640 | −32.019 | ⋯ | 0.395 | 0 | 23.285 | −2.006 | 9.631 | 6.616 | 11110001 | 1111 | 0.000 |
| 10000-155364-1109 | 0.642 | −32.057 | ⋯ | 0.351 | 2048 | 21.492 | 0.647 | 10.069 | 8.345 | 11111100 | 1111 | 0.000 |
| 10000-210677-4322 | 1.039 | −21.322 | ⋯ | 0.475 | 0 | ⋯ | 1.963 | 9.331 | 8.301 | 10111101 | 1011 | 0.000 |
| 10000-210677-4738 | 1.072 | −21.337 | ⋯ | 0.487 | 1 | ⋯ | −0.276 | 9.211 | 8.266 | 10101000 | 1011 | 0.000 |
| 10000-56544-12916 | 359.991 | −55.992 | ⋯ | 0.473 | 1 | 24.078 | 0.905 | 9.914 | 8.505 | 11111110 | 2222 | 0.250 |
| 10000-56544-12918 | 359.991 | −55.993 | ⋯ | 0.096 | 1 | 18.784 | −0.653 | 9.855 | 8.791 | 11111111 | 2222 | 0.188 |
| 10000-56544-12919 | 359.991 | −55.957 | ⋯ | 0.394 | 1 | 23.636 | 1.044 | 9.918 | 8.262 | 11111001 | 2222 | 0.203 |
| 10000-56544-12922 | 359.992 | −56.001 | ⋯ | 0.342 | 1 | 24.774 | −0.220 | 9.927 | 7.403 | 11111010 | 2222 | 0.031 |
| 10000-340449-1 | 0.244 | 1.731 | ⋯ | 0.421 | 1 | 21.349 | 0.594 | 10.109 | 8.437 | 11111100 | 2222 | 0.406 |

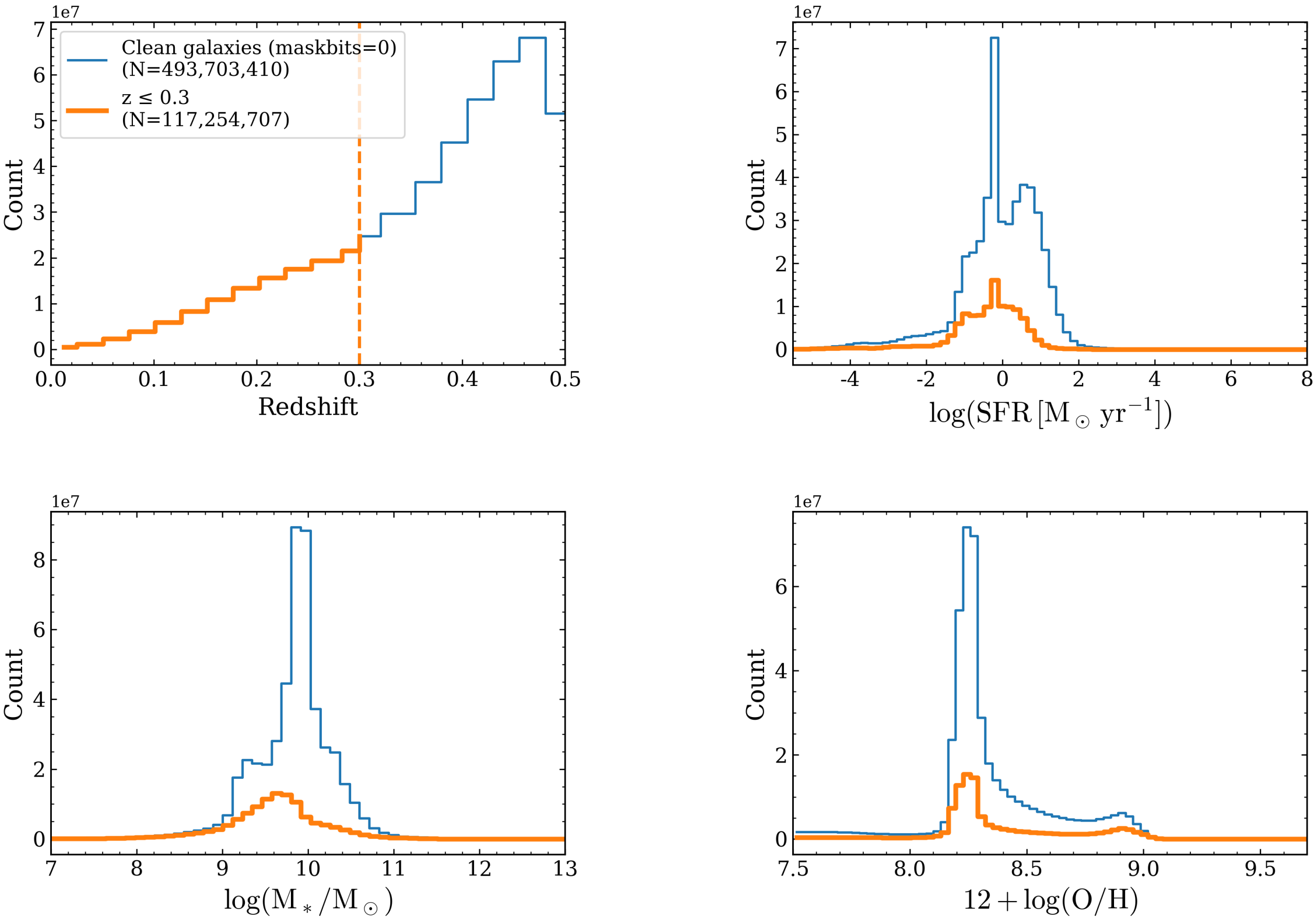

**Figure 4.** Distributions of redshift and physical properties in our properties catalog. The blue histogram indicates the clean galaxy sample (`maskbits`=0) over the full redshift range, while the orange histogram shows the corresponding subsample with $z \leqslant 0.3$.

**Table 6**
Description of the Columns in Our Physical Properties Catalog

| Name | Format | Unit | Description |
|---|---|---|---|
| desiid | string | ⋯ | Unique DESI LS target identifier, constructed from `release-brickid-objid`. |
| RA | float | deg | R.A. |
| DEC | float | deg | decl. |
| $z_{\mathrm{spec}}$ | float | ⋯ | Spectroscopic redshift collected from multiple spectroscopic surveys. |
| $z_{\mathrm{phot}}$ | float | ⋯ | Photometric redshift from C. Li et al. (2024). |
| maskbits | int | ⋯ | LS photometric quality flag; `maskbits`=0 indicates uncontaminated photometry. |
| MAG_R | float | mag | *r*-band model magnitude from DESI LS DR10. |
| log(SFR) | float | $\log(M_\odot\ \mathrm{yr}^{-1})$ | Logarithm of the star formation rate. |
| $\log(\mathrm{M}_*)$ | float | $\log(M_\odot)$ | Logarithm of the stellar mass. |
| 12 + log(O/H) | float | ⋯ | Gas-phase oxygen abundance. |
| Cat_FLAG | char(8) | ⋯ | Photometric completeness flag across bands (1: valid, 0: filled value). |
| Image_FLAG | char(4) | ⋯ | Image availability flag (1: complete cutout; 0: missing image; 2: incomplete cutout). |
| Image_Padded | float | – | Fraction of padded pixels in the image cutout (0: no padding; $0 <$ value $\leqslant 1$: partially padded; NaN: missing image). |

(iii) DESI Stellar Mass and Emission Line catalog (hereafter DESI-SMEL; H. Zou et al. 2024); (iv) DESI AGN Host Galaxies Physical Properties VAC (hereafter DESI-AGNHost; M. Siudek et al. 2024); (v) DESI PRObabilistic Value-Added Bright Galaxy Survey Catalog (hereafter DESI-PROVABGS; C. Hahn et al. 2023; K. J. Kwon et al. 2023); and (vi) GALEX–SDSS–WISE Legacy Catalog X2 (hereafter GSWLC-X2; S. Salim et al. 2016, 2018). The wavelength coverage, data modality, and SED or spectral fitting techniques differ across the catalogs. For a fair comparison, we select reliable and clean samples from each catalog based on the recommended quality flags and cross-match them against our gold sample using a 1″ matching radius. A comparison of the three galaxy properties is summarized in Table 7, using the evaluation metrics defined in Section 4.1, which here reflect consistency with other VACs rather than correctness.

Overall, our catalog shows good consistency with most reference catalogs in the low-redshift region across all three properties, supporting the reliability of our photometry-based estimates in the local Universe obtained using our multimodal deep learning model. In addition to the consistency with MPA-JHU, we also obtain broadly consistent results with GSWLC-X2, with a modest increase in scatter in SFR estimates, which is reasonable given that our model relies solely on optical-to-infrared photometry, whereas GSWLC-X2 incorporates ultraviolet photometry from Galaxy Evolution Explorer (GALEX; D. C. Martin et al. 2005), providing additional constraints on recent star formation. When compared with spectroscopic-based catalogs, including SDSS Firefly, PROVABGS, and DESI-SMEL, we observe somewhat larger offsets, as expected since our approach is based only on broadband photometric measurements and does not exploit the richer information available in galaxy spectra.

**Table 7**
Comparison of the Three Properties Estimates in Our Catalog and Reference Catalogs

| Property | Catalog | No. | MSE | RMSE | NRMSE | MAE | $\sigma$ | Bias | Outlier | NMAD | $r$ |
|---|---|---|---|---|---|---|---|---|---|---|---|
| SFR | SDSS MPA-JHU DR8 | 123,467 | 0.111 | 0.334 | 0.043 | 0.232 | 0.333 | 0.022 | 0.012 | 0.245 | 0.906 |
| ... | DESI-SMEL | 164,152 | 0.264 | 0.513 | 0.054 | 0.432 | 0.293 | 0.421 | 0.044 | 0.589 | 0.877 |
| ... | GSWLC-X2 | 114,711 | 0.152 | 0.390 | 0.060 | 0.286 | 0.371 | −0.120 | 0.012 | 0.311 | 0.890 |
| $M_*$ | SDSS MPA-JHU DR8 | 132,712 | 0.017 | 0.132 | 0.021 | 0.067 | 0.132 | 0.000 | 0.019 | 0.057 | 0.975 |
| ... | SDSS Firefly | 184,059 | 0.212 | 0.460 | 0.051 | 0.322 | 0.414 | −0.201 | 0.024 | 0.350 | 0.795 |
| ... | DESI-AGNHost | 55,060 | 0.163 | 0.403 | 0.033 | 0.299 | 0.360 | 0.182 | 0.015 | 0.363 | 0.894 |
| ... | DESI-SMEL | 164,152 | 0.079 | 0.281 | 0.029 | 0.193 | 0.256 | −0.117 | 0.015 | 0.216 | 0.914 |
| ... | DESI-PROVABGS | 45,878 | 0.207 | 0.455 | 0.070 | 0.359 | 0.316 | −0.328 | 0.045 | 0.441 | 0.883 |
| ... | GSWLC-X2 | 114,711 | 0.019 | 0.137 | 0.028 | 0.103 | 0.103 | −0.090 | 0.018 | 0.129 | 0.984 |
| 12+log(O/H) | SDSS MPA-JHU DR8 | 34,444 | 0.009 | 0.094 | 0.060 | 0.066 | 0.091 | −0.021 | 0.011 | 0.078 | 0.884 |

In addition, intrinsic differences among these catalogs are primarily due to the use of different SEDs or spectral fitting codes (e.g., CIGALE, FIREFLY, and FASTSPECFIT; J. Moustakas et al. 2023), as well as different modeling assumptions, including choices of IMFs and stellar population libraries, which inevitably lead to systematic offsets in cross-catalog comparisons. Across different physical properties, SFR generally exhibits larger scatter than stellar mass when compared with other catalogs, consistent with the discussion in Section 4.2.1 and reflecting the intrinsically greater difficulty of estimating SFR.

Nevertheless, the overall agreement with commonly used reference catalogs supports the robustness and reliability of our catalog, providing a useful and complementary data resource for downstream scientific analyses.

#### *5.3.2. Exploration of Galaxy Scaling Relations*

In this section, we explore scaling relations among the derived galaxy physical properties, focusing on the star-forming main sequence and the MZR. Rather than performing a detailed quantitative analysis, we present a qualitative examination of these relations to assess the overall physical plausibility of the catalog and to illustrate its potential for future scientific applications.

The distribution of galaxies in the log(SFR)-log($M_*$) plane provides a global view of star formation activity across the galaxy population. Star-forming galaxies are known to populate a relatively tight sequence in this plane, commonly referred to as the star-forming main sequence (MS). This sequence has been established as one of the most fundamental scaling relations in galaxy evolution, holding from the local Universe up to high redshift ($z \sim 6$) (J. Brinchmann et al. 2004; D. Elbaz et al. 2007; J. S. Speagle et al. 2014; C. Schreiber et al. 2015). The position of a galaxy relative to the MS is widely used to distinguish normal star-forming galaxies from starburst galaxies above the sequence and quiescent galaxies below it, making the MS a key diagnostic for galaxy evolutionary stages (G. Rodighiero et al. 2011; M. Troncoso et al. 2025).

Figure 5 (left) presents the distribution of $M_*$ and SFR for our gold sample. A clear bimodal structure is visible, consisting of a prominent star-forming sequence and a population of low-SFR galaxies consistent with quiescent galaxies. This overall structure is in good agreement with the general picture of galaxy populations in the local Universe (J. Brinchmann et al. 2004; A. Saintonge et al. 2016).

We further cross-match our sample with the AGN/QSO Value-Added Catalog for DESI DR1 (hereafter DESI-AGNQSO; Juneau et al. 2026, in preparation) within a 1″ radius, which provides independent classifications of star-forming galaxies and AGN based on spectroscopic diagnostics. In this work, we adopt the classifications based on the [N II] BPT diagram (J. A. Baldwin et al. 1981; L. J. Kewley et al. 2001; G. Kauffmann et al. 2003; K. Schawinski et al. 2007). The 95% density contours of the star-forming galaxies and AGN population are shown as cyan and red, respectively. We further compute the median SFR of star-forming galaxies in bins of stellar mass with a typical bin width of ∼0.2 dex. The resulting relation, shown as a cyan line, exhibits a tight correlation between SFR and stellar mass. We further overlay the local MS from D. Elbaz et al. (2007) with a typical ±0.3 dex scatter, which is derived from SDSS galaxies. The relation of the star-forming population in our catalog closely follows this reference relation, showing good agreement in both slope and normalization. Meanwhile, AGN-host galaxies are preferentially distributed toward the massive end of the sequence and extend into the intermediate region between the star-forming and quiescent populations, commonly referred to as the Green Valley (K. Nandra et al. 2007; T. K. Wyder et al. 2007).

The agreement with both independent galaxy classifications and established MS suggests that our catalog captures the key structures of galaxy populations in the local Universe, and enables the selection of distinct galaxy populations, such as star-forming, transitional, and quiescent systems, for statistical studies of galaxy evolution.

We next explore the MZR (e.g., J. Lequeux et al. 1979; C. A. Tremonti et al. 2004), another fundamental scaling relations in galaxy evolution. The MZR describes the correlation between stellar mass and gas-phase metallicity, whereby more-massive galaxies tend to exhibit higher chemical abundances. This relation reflects the combined effects of star formation, gas inflows, and metal-enriched outflows, which regulate the chemical enrichment of galaxies over cosmic time (e.g., D. R. Garnett 2002; S. L. Ellison et al. 2008).

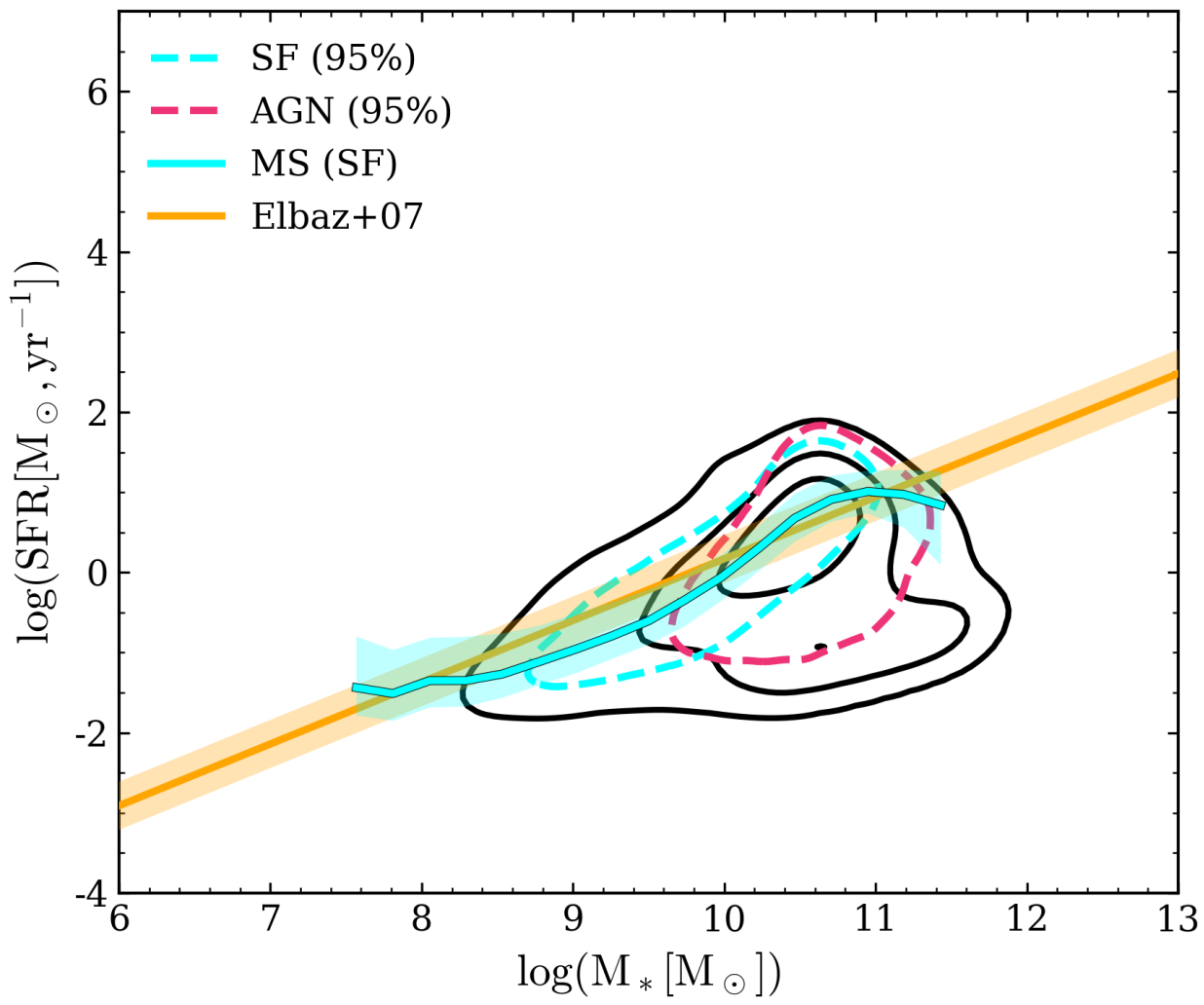

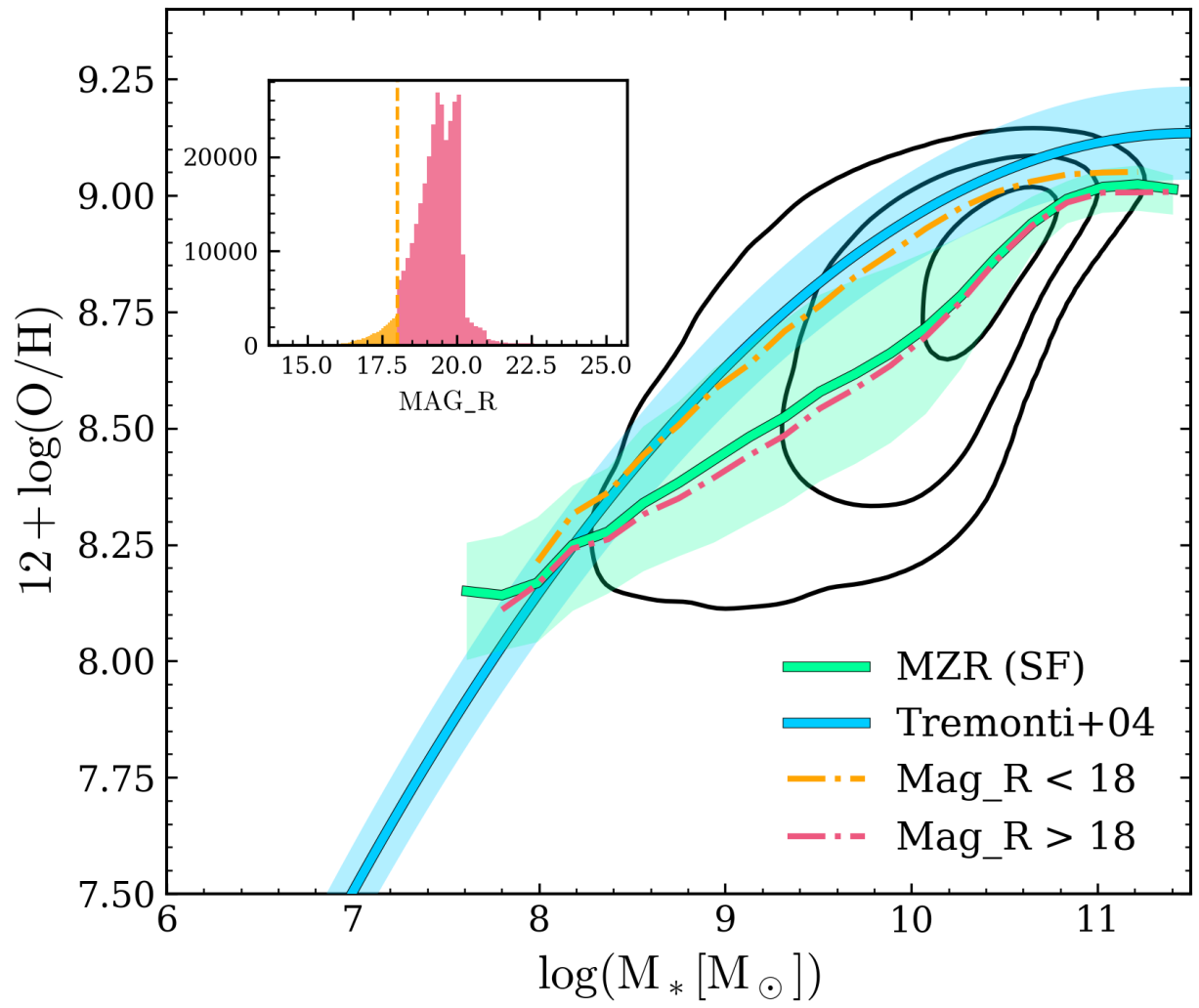

**Figure 5.** Left panel: distribution of the gold sample from our catalog in the SFR–$M_*$ plane. The black contours indicate the $1\sigma$, $2\sigma$, and $3\sigma$ density levels of the full sample. The cyan and red contours show the 95% density regions of star-forming galaxies and AGNs, respectively, as classified in the DESI-AGNQSO VAC. The cyan line shows the median MS of our star-forming galaxies, with the shaded region representing the 16th–84th percentile range. The orange line represents the local MS from D. Elbaz et al. (2007), with the shaded region indicating a typical $\pm 0.3$ dex scatter. Right panel: distribution of the star-forming galaxy subsample selected via cross-matching with the DESI-AGNQSO VAC in the $12 + \log(\mathrm{O/H})$–$M_*$ plane. The black contours indicate the $1\sigma$, $2\sigma$, and $3\sigma$ density levels, as in the left panel. The green line shows the median MZR of our sample, with the shaded region representing the 16th–84th percentile range. The blue line denotes the local MZR from C. A. Tremonti et al. (2004), with the shaded region indicating the reported $\pm 0.1$ dex scatter. The orange and pink dotted–dashed lines represent the MZRs of subsamples with MAG_R $< 18$ and MAG_R $> 18$, respectively. The inset panel (top left) shows the distribution of MAG_R for the star-forming subsample, with the vertical orange line marking the threshold at MAG_R $= 18$.

Figure 5 (right panel) shows the distribution of $12 + \log(\mathrm{O/H})$ as a function of stellar mass for the star-forming galaxy subsample in our gold sample, selected via cross-matching with the DESI-AGNQSO VAC, consistent with the star-forming sample used in the MS analysis. We overplot the local MZR derived by C. A. Tremonti et al. (2004, hereafter T04), together with its reported $\pm 0.1$ dex scatter, which was calibrated using star-forming galaxies from the SDSS.

Overall, our sample follows the general trend of the established MZR, exhibiting a rising metallicity at low stellar masses and a gradual flattening toward the high-mass regime. Although the $1\sigma$ uncertainty of our relation overlaps with that of T04, a systematic offset is evident, with our MZR shifted toward lower metallicities.

This discrepancy is likely related to differences in the galaxy samples. T04 selected star-forming galaxies from the SDSS with Petrosian $r$-band magnitudes in the range $14.5 < r < 17.77$. To investigate the impact of photometric depth, we examine the distribution of $r$-band model magnitudes in our sample and divide it into two subsamples using a threshold of MAG_R $= 18$, ignoring the minor differences between Petrosian and model magnitudes. As shown in Figure 5 (right panel), the MZR derived from the brighter subsample (MAG_R $< 18$) is consistent with the T04 relation, while the fainter subsample (MAG_R $> 18$) closely follows the overall MZR.

This behavior suggests that the observed offset is primarily driven by differences in photometric depth between SDSS and DESI LS, with the latter probing a fainter galaxy population (M. A. Strauss et al. 2002; A. Dey et al. 2019). The lower metallicities observed in the fainter subsample are also consistent with the luminosity–metallicity relation, in which less-luminous galaxies tend to have lower chemical abundances (see Figure 4 and Equation (2) in T04). In addition, differences in redshift coverage may also contribute to the discrepancy. While the T04 sample is limited to $z \lesssim 0.25$, our sample extends to $z \sim 0.5$. However, since magnitude and redshift are intrinsically coupled in magnitude-limited surveys, disentangling their individual effects is nontrivial. In this work, we therefore consider the magnitude dependence to be the dominant factor.

The presence of both the MS and the MZR in our sample suggests that the derived galaxy properties broadly capture the expected physical correlations, highlighting the potential of this catalog for future studies of galaxy populations and their evolution. We emphasize that this analysis is primarily qualitative, and a more rigorous quantitative investigation, accounting for various sources of systematic uncertainty, is beyond the scope of this work.

### 5.4. Caveats and Limitations

Despite the large scale and broad applicability of the physical properties catalog presented in this work, several important limitations and sources of systematic uncertainty should be clearly acknowledged.

First, as the model was trained and validated primarily on galaxies with redshifts predominantly below 0.3, its predictions are expected to be most reliable within this range, while results for galaxies up to $z \sim 0.5$ are also included to extend the analysis. This limitation implies that the current catalog cannot be directly used for studies of the intermediate- or high-redshift Universe. In future work, we plan to extend the training dataset to higher-redshift galaxies, which will allow the model to provide reliable estimates across a broader redshift range.

Second, the estimated physical properties are derived from a supervised, data-driven model trained on the MPA-JHU DR8 catalog. As a result, the model learns statistical mappings that are valid primarily within the distribution and selection function of the training sample. In particular, gas-phase metallicities in the training data are available only for star-forming galaxies, implying that the metallicity estimates in this catalog should not be interpreted as physically meaningful for quiescent or AGN-dominated systems. Users are therefore strongly encouraged to restrict metallicity-based analyses to samples consistent with the star-forming population or to apply appropriate classification and quality cuts.

Third, redshift information enters the model as an explicit input feature, and uncertainties in redshift—especially for objects with photometric redshifts—inevitably propagate into the estimated properties. While the model captures the average trends present in the training data, it does not explicitly propagate redshift uncertainties on an object-by-object basis. Consequently, estimates for galaxies relying on photometric redshifts should be interpreted in a statistical sense, and analyses focusing on individual objects or small subsamples should be treated with caution.

Fourth, the handling of missing or low-quality photometric measurements represents another source of potential systematic bias. Although a unified filling strategy is adopted to ensure numerical stability during training and subsequent application, extreme patterns of missing data or severe photometric contamination may lead to biased estimates for a subset of objects. The provided quality flags are therefore an essential component of the catalog, and users are advised to apply conservative selections when precision is required.

More generally, the physical quantities provided in this catalog are not intended to replace spectroscopic or physically motivated modeling for individual galaxies. Instead, they are optimized for large-scale statistical studies, where systematic uncertainties at the level of individual objects average out, and robust population-level trends can be recovered. When used within these recommended boundaries, the catalog offers a powerful and efficient resource for exploring galaxy demographics across the low-redshift Universe.

## 6. Conclusion

In this work, we have presented a large-scale catalog of galaxy physical properties constructed for the DESI LS DR10. Using a multimodal deep learning framework that jointly exploits optical imaging data and photometric catalog features, we estimate three fundamental galaxy properties—SFR, stellar mass, and gas-phase oxygen abundance—for 547,656,694 galaxies at low redshift ($z \leqslant 0.5$).

The proposed framework combines a ResNet-based CNN to extract spatial and morphological information from multiband optical images with an MLP that processes catalog-level photometric features. By integrating these complementary data modalities, the model captures both structural and broadband photometric information relevant to galaxy physical properties. Trained on reference measurements from the MPA-JHU DR8 catalog, the model achieves competitive performance compared to existing photometry-based and image-based approaches, while enabling efficient estimation at the scale required by modern imaging surveys.

We validated the resulting catalog through multiple complementary tests. These include direct comparisons with independent value-added catalogs based on spectroscopic and SED-fitting techniques, as well as the exploration of key galaxy scaling relations such as the star-forming main sequence and the stellar MZR in the local Universe. The consistency observed in these comparisons demonstrates that the catalog preserves the dominant astrophysical information necessary for statistical studies of galaxy populations.

It is important to emphasize that the physical properties provided in this catalog are optimized for population-level and statistical analyses rather than precision measurements of individual galaxies. The catalog provides point estimates only and does not include per-object uncertainties, and the estimates are subject to systematic uncertainties associated with the training labels, photometric redshift errors, and data incompleteness, which are explicitly documented through quality flags and usage recommendations. When used within these defined boundaries, the catalog offers a robust and homogeneous resource for large-sample investigations of galaxy evolution. Users are therefore encouraged to rely on ensemble statistics and apply additional quality cuts when performing analyses sensitive to individual measurements.

By providing uniform estimates of SFR, stellar mass, and metallicity for the low-redshift DESI LS DR10 galaxy sample, this catalog substantially expands the scientific utility of the Legacy Surveys. It enables a wide range of applications, including studies of galaxy demographics, environmental trends, target selection for spectroscopic follow-up, and cross-survey comparisons with current and future datasets. More broadly, this work demonstrates the potential of multimodal, data-driven approaches for transforming large photometric surveys into physically informative datasets, paving the way for similar efforts in upcoming surveys such as DESI extensions and LSST.

## Acknowledgments

We sincerely thank the reviewer for the time and valuable feedback, which has helped us improve the manuscript. This work is supported by National Natural Science Foundation of China (NSFC; grant Nos. 12373110, 12273076, 12403102, 12103070, 12273077, and 12133001), Strategic Priority Research Program of the Chinese Academy of Sciences (XDB0550101). Data resources are supported by China National Astronomical Data Center (NADC) and Chinese Virtual Observatory (China-VO). This work is supported by Astronomical Big Data Joint Research Center, co-founded by National Astronomical Observatories, Chinese Academy of Sciences, and Alibaba Cloud.

This research used data obtained with the Dark Energy Spectroscopic Instrument (DESI). DESI construction and operations are managed by the Lawrence Berkeley National Laboratory. This material is based upon work supported by the U.S. Department of Energy, Office of Science, Office of High-Energy Physics, under contract No. DE-AC02-05CH11231, and by the National Energy Research Scientific Computing Center, a DOE Office of Science User Facility under the same contract. Additional support for DESI was provided by the U.S. National Science Foundation (NSF), Division of Astronomical Sciences under contract No. AST-0950945 to the NSF's National Optical-Infrared Astronomy Research Laboratory; the Science and Technology Facilities Council of the United Kingdom; the Gordon and Betty Moore Foundation; the Heising-Simons Foundation; the French Alternative Energies and Atomic Energy Commission (CEA); the National Council of Humanities, Science and Technology of Mexico (CONAHCYT); the Ministry of Science and Innovation of Spain (MICINN), and by the DESI Member Institutions: www.desi.lbl.gov/collaborating-institutions. The DESI collaboration is honored to be permitted to conduct scientific research on I'oligam Du'ag (Kitt Peak), a mountain with particular significance to the Tohono O'odham Nation. Any opinions, findings, and conclusions or recommendations expressed in this material are those of the author(s) and do not necessarily reflect the views of the U.S. National Science Foundation, the U.S. Department of Energy, or any of the listed funding agencies.

Funding for the Sloan Digital Sky Survey IV has been provided by the Alfred P. Sloan Foundation, the U.S. Department of Energy Office of Science, and the Participating Institutions.

SDSS-IV acknowledges support and resources from the Center for High Performance Computing at the University of Utah. The SDSS website is www.sdss4.org.

SDSS-IV is managed by the Astrophysical Research Consortium for the Participating Institutions of the SDSS Collaboration including the Brazilian Participation Group, the Carnegie Institution for Science, Carnegie Mellon University, Center for Astrophysics | Harvard & Smithsonian, the Chilean Participation Group, the French Participation Group, Instituto de Astrofísica de Canarias, The Johns Hopkins University, Kavli Institute for the Physics and Mathematics of the Universe (IPMU)/University of Tokyo, the Korean Participation Group, Lawrence Berkeley National Laboratory, Leibniz Institut für Astrophysik Potsdam (AIP), Max-Planck-Institut für Astronomie (MPIA Heidelberg), Max-Planck-Institut für Astrophysik (MPA Garching), Max-Planck-Institut für Extraterrestrische Physik (MPE), National Astronomical Observatories of China, New Mexico State University, New York University, University of Notre Dame, Observatário Nacional/MCTI, The Ohio State University, Pennsylvania State University, Shanghai Astronomical Observatory, United Kingdom Participation Group, Universidad Nacional Autónoma de México, University of Arizona, University of Colorado Boulder, University of Oxford, University of Portsmouth, University of Utah, University of Virginia, University of Washington, University of Wisconsin, Vanderbilt University, and Yale University.

*Software:* NumPy (C. R. Harris et al. 2020), Astropy (Astropy Collaboration et al. 2013, 2018, 2022), Matplotlib (J. D. Hunter 2007), TOPCAT (M. B. Taylor 2005), PyTorch (A. Paszke et al. 2019), Weights & Biases (https://wandb.ai/site/experiment-tracking/).

## Data Availability

The galaxy physical properties catalog in this work is publicly available at doi:10.12149/101777. We additionally provide an interactive, user-friendly web service for querying the catalog at https://nadc.china-vo.org/ai/query/legacy_galaxy_physical_parameters/f. The code for model training and evaluation is publicly available at https://github.com/RuiNov1st/multimodal-desilsdr10-properties-vac.

The DESI Legacy Imaging Surveys DR10 data products are publicly available at https://portal.nersc.gov/cfs/cosmo/data/legacysurvey/dr10/. The photometric redshift catalog for DR10 is available at the National Astronomical Data Center (C. Li 2024).

The reference catalogs used for cross-catalog comparisons in this work are publicly available. The SDSS MPA-JHU DR8 galaxy properties catalog can be accessed at https://www.sdss4.org/dr17/spectro/galaxy_mpajhu/. The eBOSS Firefly value-added catalog (DR16) is available at https://www.sdss4.org/dr16/spectro/eboss-firefly-value-added-catalog/. The DESI DR1 value-added catalogs used in this work, including the Stellar Mass and Emission Line catalog, the AGN Host Galaxies Physical Properties catalog, the PROVABGS catalog, and the AGN/QSO Value-Added Catalog, are available through the DESI data release portal at https://data.desi.lbl.gov/doc/releases/. The GALEX–SDSS–WISE Legacy Catalog X2 (GSWLC-X2) is publicly available at https://salims.pages.iu.edu/gswlc/.

## Appendix A
## Data Features

In Section 2.4, we introduced the catalog and imaging features constructed as inputs for the model. In this appendix section, we provide a detailed description of the construction and selection of catalog features, as well as additional experiments to validate the effectiveness of incorporating color information into the imaging inputs.

### *A.1. Catalog Data Features*

The DESI LS DR10 photometric catalog provides comprehensive measurements of galaxy light distributions, including model magnitudes and aperture magnitudes measured within multiple aperture radii across all bands. These measurements, together with their combinations, encode both photometric and spatial information that is closely linked to galaxy physical properties. To fully exploit the available information, we construct a total of 1626 photometric features by combining magnitude measurements, computing magnitude differences, and incorporating auxiliary information including Galactic extinction $E(B - V)$ and redshift. Specifically, the 1624 magnitude-related features include:

(1) Absolute photometric measurements, including model magnitudes and aperture magnitudes in all bands and aperture radii, which describe the overall brightness of galaxies (e.g., MAG_G, MAG_R, APMAG_G_1), 60 in total.

(2) Colors, defined as magnitude differences between different photometric bands using model or aperture magnitudes, including combinations across different aperture radii (e.g., MAG_G – MAG_R, MAG_G – MAG_I, APMAG_G_1 – APMAG_R_1, APMAG_G_1 – APMAG_R_2), 1360 in total.

(3) Radial light distributions, constructed from differences between model magnitudes and aperture magnitudes or between aperture magnitudes at different radii, characterizing the spatial distribution of galaxy light (e.g., MAG_G – APMAG_G_1, APMAG_G_1 – APMAG_G_2), 204 in total.

To identify the most informative catalog features for different galaxy property predictions, we perform feature selection based on the feature importance scores provided by CatBoost. Feature importance scores derived from the CatBoost model quantify the contribution of each feature to the predictive performance, allowing less informative or redundant features to be removed while retaining the most relevant inputs. Specifically, we train a CatBoost model using only catalog features on the validation set, incorporating all 1626 constructed features, and obtain an importance score for each feature. To remove the redundant and noisy features, we further adopt a threshold-based selection strategy, in which features with importance scores above a given threshold are retained. By varying the threshold and evaluating the corresponding validation performance, we identify feature subsets that achieve a good balance between computational efficiency and estimation accuracy. We finally select 28 features for SFR, 50 features for stellar mass, while all 1626 features are retained for $12 + \log(\mathrm{O/H})$. The varying number of selected features across different galaxy properties may indicate their distinct information requirements. SFR and stellar mass can be constrained by a compact set of photometric features tracing luminosity, color, and large-scale light distribution, whereas gas-phase oxygen abundance is intrinsically spectroscopic and benefits from a more comprehensive feature set under photometry-only conditions.

The 50 selected features used for the stellar mass estimation are listed below, while the feature sets adopted for SFR and $12 + \log(\mathrm{O/H})$ are provided with our published VAC:

MAG_G − MAG_R, MAG_R − MAG_Z, MAG_G, MAG_R − MAG_W1, MAG_R, MAG_R − MAG_I, MAG_W1 − MAG_W2, APMAG_R_1 − APMAG_R_2, MAG_I − MAG_Z, MAG_Z − MAG_W1, APMAG_G_1 − APMAG_G_2, MAG_Z, APMAG_Z_2 −APMAG_W1_2, APMAG_Z_5 − APMAG_W1_5, MAG_G − APMAG_G_1, MAG_R − MAG_W2, MAG_G − MAG_I, APMAG_G_5 − APMAG_R_5, APMAG_W1_2 − APMAG_W1_3, MAG_G − MAG_Z, APMAG_R_2 − APMAG_R_3, APMAG_Z_2 − APMAG_Z_3, APMAG_R_1 − APMAG_I_1, MAG_W1, APMAG_G_1 − APMAG_R_1, APMAG_R_3 − APMAG_R_4, APMAG_Z_4 − APMAG_W1_4, MAG_I − APMAG_I_1, APMAG_G_4 − APMAG_R_4, MAG_G − MAG_W2, APMAG_Z_1 − APMAG_Z_2, APMAG_G_5 − APMAG_G_6, APMAG_Z_5 − APMAG_Z_6, APMAG_I_5 −APMAG_Z_5, APMAG_Z_3 − APMAG_Z_4, MAG_G − MAG_W1, MAG_W1 − MAG_W2, APMAG_G_6 − APMAG_R_6, APMAG_G_4 − APMAG_G_5, MAG_R− APMAG_R_1, APMAG_G_6 − APMAG_G_7, MAG_I −MAG_W1, APMAG_Z_4 − APMAG_Z_5, APMAG_G_7 − APMAG_R_7, MAG_I, APMAG_W1_3 − APMAG_W1_4, APMAG_R_5 − APMAG_I_5, APMAG_Z_3 − APMAG_W1_3, $E(B-V)$, redshift.

### A.2. Imaging Data Features

To exploit the color information encoded in the photometric images and avoid requiring the model to learn such information implicitly from the fluxes, we explicitly construct color channels by computing $g - r$, $r - i$, and $i - z$, and concatenate them with the original $g$, $r$, $i$, and $z$ bands. To validate the effectiveness of this design, we conduct a comparative experiment using image inputs with and without color information on the same test set. The results are presented in Table 8. We find that incorporating explicit color features consistently enhances model performance. This suggests that, although color information can be inferred from photometric bands, providing it explicitly allows the model to more effectively capture features related to stellar populations and SFH. Consequently, color information proves to be a valuable and nonredundant input in practice.

**Table 8**
Comparison of Model Performance with and without Color Information in the Image Inputs across Different Galaxy Properties

| Property | Feature | MSE | RMSE | NRMSE | MAE | $\sigma$ | Bias | Outlier | NMAD | $r$ |
|---|---|---|---|---|---|---|---|---|---|---|
| SFR | Image (without color) | 0.195 | 0.441 | 0.056 | 0.324 | 0.440 | −0.037 | 0.012 | 0.366 | 0.829 |
| SFR | Image (with color) | 0.174 | 0.418 | 0.053 | 0.304 | 0.416 | −0.032 | 0.012 | 0.338 | 0.848 |
| $M_*$ | Image (without color) | 0.054 | 0.233 | 0.036 | 0.152 | 0.232 | −0.014 | 0.012 | 0.153 | 0.928 |
| $M_*$ | Image (with color) | 0.033 | 0.181 | 0.028 | 0.116 | 0.180 | −0.019 | 0.015 | 0.110 | 0.958 |
| $12 + \log(\mathrm{O/H})$ | Image (without color) | 0.011 | 0.106 | 0.066 | 0.074 | 0.106 | −0.003 | 0.015 | 0.079 | 0.878 |
| $12 + \log(\mathrm{O/H})$ | Image (with color) | 0.010 | 0.010 | 0.061 | 0.071 | 0.099 | −0.014 | 0.016 | 0.077 | 0.893 |

# Appendix B
# Quality-related Factors in the Catalog

In Sections 2 and 5.1, we have introduced several factors that may affect the performance of the property estimation model and the quality of our catalog, including missing values in the photometric catalog, incomplete image cutouts, and the use of photometric redshifts as input features. In this section, we quantitatively evaluate how these factors affect the estimated physical properties.

### B.1. Missing Values in the Photometric Catalog

Missing photometric measurements are encoded by the `Cat_FLAG`, an 8-bit integer indicating which bands are unavailable and subsequently filled with predefined maximum values. To assess the impact of missing photometric information on model performance, we use the training sample with reference labels from the MPA-JHU DR8 catalog and group galaxies according to the number of missing photometric bands, since such missing values are already present in the training sample.

For each group, we compute an evaluation metric scatter $\sigma$ and examine its dependence on the number of missing bands. The results are shown in the left panel of Figure 6, where the background histogram indicates the distribution of sources in each missing-band bin.

We find that the scatter $\sigma$ remains stable for galaxies with up to three missing bands and shows a noticeable increase only when four bands are missing. The background distribution reveals that the majority of sources have between zero and three missing bands, while objects with four or five missing bands are rare. The apparent decrease in $\sigma$ for the five-missing-band bin is driven by the very small sample size and therefore is not statistically meaningful.

Overall, these results suggest that our catalog remains reliable in the presence of missing photometric values, as long as only a small number of bands are missing. In practice, the impact depends on both the total number of missing bands and which specific bands are absent, and should therefore be evaluated on a case-by-case basis.

### B.2. Image Incompleteness

For incomplete image cutouts, the `Image_FLAG` and `Image_Padded` columns characterize the integrity of the imaging data by indicating whether an image has been padded and quantifying the fraction of padded regions, respectively. To assess the impact of image incompleteness on model performance, we conduct a set of simulation-based experiments using 10,000 randomly selected clean galaxies. Here, "clean" samples refer to galaxies with complete catalog measurements and intact image cutouts, without any artificial masking or filling. Starting from these clean images, we simulate edge-cut effects by padding 25%–75% of the image area with zeros. The padded region is randomly applied to one of the four directions (top, bottom, left, or right), while all catalog features are kept unchanged.

Figure 6 (right panel) shows the model performance as a function of the padded image fraction, quantified by the scatter metric $\sigma$. We find that padding up to 25% of the image area has a negligible impact on the derived physical properties. However, as the padded fraction increases, the estimation accuracy degrades rapidly, as expected when a substantial fraction of the image information is missing.

Overall, this simulation-based experiment demonstrates that image padding can affect model performance when a large fraction of the image is missing. However, according to Section 5.2, only a small fraction of the images in our catalog is incomplete, and the mean padded fraction is below 25%. Therefore, the image data for the majority of the samples is still being considered to be reliable.

### B.3. Photometric Redshift Substitution

We further examine the impact of replacing spectroscopic redshift ($z_{\rm spec}$) with photometric redshift ($z_{\rm phot}$) from C. Li et al. (2024) as an input feature to the model. We randomly select 10,000 galaxies from the training sample, for which $z_{\rm spec}$ from the MPA-JHU catalog are available, and cross-match them with the photometric redshift catalog to obtain the corresponding $z_{\rm phot}$ measurements. For these 10,000 galaxies, the bias and NMAD between $z_{\rm phot}$ and $z_{\rm spec}$ are 0.00008 and 0.034, respectively, with an outlier fraction of 0.2%. We then replace the $z_{spec}$ with the $z_{phot}$ in the input catalog and compare the resulting property estimates with those obtained using $z_{spec}$. The results are summarized in Table 9.

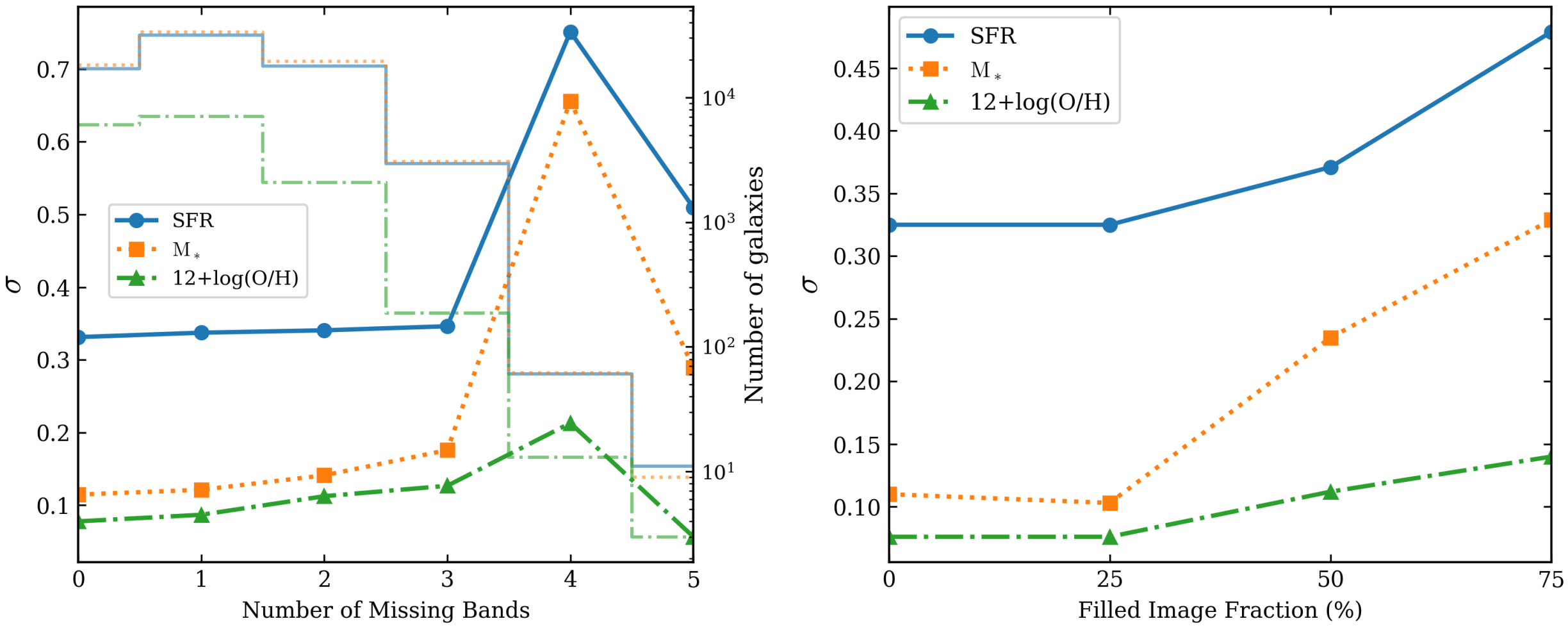

**Figure 6.** Impact of catalog incompleteness and image padding on model performance. Left panel: the scatter metric $\sigma$ as a function of the number of missing photometric bands in the catalog for SFR, stellar mass ($M_*$), and oxygen abundance (12+log(O/H)). The step curves in the background indicate the corresponding number of galaxies in each bin of missing photometric bands for the three properties. Right panel: the scatter metric $\sigma$ as a function of the filled image fraction, based on simulated image-padding experiments.

**Table 9**
Comparison of Model Performance Using Spectroscopic ($z_{spec}$) and Photometric ($z_{phot}$) Redshifts as Input Features, Evaluated on the Same Subsample

| Property | $z_{spec}$ | | | | | $z_{phot}$ | | | | |
|---|---|---|---|---|---|---|---|---|---|---|
| | NRMSE | $\sigma$ | Bias | Outlier | NMAD | NRMSE | $\sigma$ | Bias | Outlier | NMAD |
| SFR | 0.053 | 0.416 | −0.032 | 0.012 | 0.338 | 0.070 | 0.357 | −0.028 | 0.010 | 0.317 |
| $M_\star$ | 0.019 | 0.106 | −0.003 | 0.017 | 0.058 | 0.065 | 0.366 | −0.006 | 0.007 | 0.346 |
| 12+log(O/H) | 0.056 | 0.076 | −0.004 | 0.015 | 0.064 | 0.056 | 0.076 | −0.004 | 0.015 | 0.064 |

We find that for SFR and oxygen abundance, substituting $z_{spec}$ with $z_{phot}$ leads to no pronounced degradation in performance. In particular, for oxygen abundance, the differences in the evaluation metrics are minimal and only appear at the level of the last few decimal places, which are not shown in the table. In contrast, stellar mass estimates exhibit a higher sensitivity to the redshift input. At the level of redshift differences considered in this experiment, replacing $z_{spec}$ with $z_{phot}$ introduces measurable changes in the derived stellar masses. This behavior is consistent with the well-known dependence of stellar mass estimation on accurate distance information.

Overall, Table 9 shows that physical property estimates based on photometric redshifts can achieve statistical accuracy comparable to those based on spectroscopic redshifts for SFR and oxygen abundance. For stellar mass, predictions corresponding to large discrepancies between $z_{spec}$ and $z_{phot}$ should be treated with caution.

## ORCID iDs

Shirui Wei https://orcid.org/0009-0005-3959-6547
Changhua Li https://orcid.org/0000-0002-6830-0013
Yanxia Zhang https://orcid.org/0000-0002-6610-5265
Chenzhou Cui https://orcid.org/0000-0002-7456-1826
Wujun Shao https://orcid.org/0009-0002-2492-6054
Zihan Kang https://orcid.org/0000-0003-3778-3566